\documentclass[prd,showpacs,letterpaper,10pt,aps,notitlepage,nofootinbib,groupedaddress]{revtex4-2}
\usepackage{changepage}
\usepackage[utf8]{inputenc}
\usepackage{amsfonts}
\usepackage{amsmath}
\usepackage{graphicx}
\usepackage{hyperref}
\usepackage{placeins}
\usepackage{mathrsfs}
\usepackage{bm}
\usepackage{multirow}
\usepackage{slashed}
\usepackage[export]{adjustbox}
\hypersetup{colorlinks, linkcolor = [rgb]{0,0.0,0.75}, citecolor = [rgb]{0,0.0,0.75}, urlcolor = [rgb]{0,0.0,0.75}}
\renewcommand{\hat}{\widehat}
\renewcommand{\tilde}{\widetilde}

\newcommand{\nb}{\phantom{0}}
\newcommand{\wm}{\phantom{-}}

\begin{document}
\title{\texorpdfstring{$\bm{\Xi_b \to \Xi}$}{Xib to Xi} form factors from lattice QCD and \texorpdfstring{\\}{} Standard-Model predictions for \texorpdfstring{$\bm{\Xi_b \to \Xi \mu^+\mu^-}$}{Xib to Xi mu(+) mu(-)} and \texorpdfstring{$\bm{\Xi_b \to \Xi \gamma}$}{Xib to Xi gamma} decays}

\author{Callum Farrell}
\author{Stefan Meinel}
\affiliation{Department of Physics, University of Arizona, Tucson, AZ 85721, USA}

\begin{abstract}
We present the first lattice QCD determination of the $\Xi_b \to \Xi$ vector, axial-vector, and tensor form factors, which are relevant for the theory of rare decays including $\Xi_b \to \Xi \ell^+\ell^-$ and $\Xi_b \to \Xi \gamma$. The calculation is performed with 2+1 flavors of domain-wall fermions at three different lattice spacings and pion masses in the range from approximately 430 to 230 MeV. The bottom quark is implemented using an anisotropic clover action. Three-point functions with a wide range of source-sink separations and model averaging are used to extract the ground-state contributions. We fit the dependence of the form factors on the momentum transfer, the pion mass, and the lattice spacing using modified $z$ expansions that account for subthreshold branch cuts, and apply dispersive bounds and asymptotic-behavior constraints to achieve controlled uncertainties in the full semileptonic kinematic region. Using our form factor results, we present Standard-Model predictions for the $\Xi_b^- \to \Xi^- \gamma$ and $\Xi_b^- \to \Xi^- \mu^+\mu^-$ branching fractions and two angular observables.
\end{abstract}

\maketitle

\FloatBarrier
\section{Introduction}
\FloatBarrier
\label{sec:intro}
Rare decays of $b$-hadrons have long been a focus of theoretical and experimental scrutiny in the search for physics beyond the Standard Model. Their persistent study has uncovered a variety of tensions between theoretical predictions and experimental measurements, often collectively referred to as the $B$ anomalies (see, e.g., Ref.~\cite{Capdevila:2023yhq}). Even though hints for lepton-flavor universality violation in $b\to s\ell^+\ell^-$ decays have disappeared after new measurements \cite{LHCb:2022vje,LHCb:2022qnv}, tensions with Standard-Model predictions persist in the branching fractions of $B \to K^{(*)} \mu^ + \mu^-$ and $B_s \to \phi \mu^ + \mu^-$ and in angular observables in the $B \to K^{*} \mu^+ \mu^- $ transitions \cite{Descotes-Genon:2013wba,Horgan:2013hoa,Horgan:2013pva,Altmannshofer:2014rta,LHCb:2014cxe,Bharucha:2015bzk,Bailey:2015dka,Horgan:2015vla,LHCb:2016ykl,Gubernari:2018wyi,LHCb:2020gog,LHCb:2021zwz,Gubernari:2022hxn,Parrott:2022rgu,Parrott:2022zte,LHCb:2023gel,Gubernari:2023puw,CMS:2024syx,CMS:2024atz,LHCb:2025mqb,LHCb:2026suh}.
Recent global fits \cite{Hurth:2025vfx,Alguero:2023jeh,Capdevila:2023yhq,Hurth:2023jwr,Wen:2023pfq,Greljo:2022jac,Gubernari:2022hxn} favor a significant shift solely in the Wilson coefficient $C_9$. To help discriminate the source of this shift from new physics or yet underestimated Standard-Model processes ~\cite{Khodjamirian:2010vf,Jager:2012uw,Jager:2014rwa,Lyon:2014hpa,Ciuchini:2015qxb,Bobeth:2017vxj,Gubernari:2020eft,Ciuchini:2022wbq,Feldmann:2023plv,Isidori:2024lng,Bordone:2024hui,Hurth:2025neo,Isidori:2025dkp}
 it is useful to study $b \to s \ell \ell $ transitions in several distinct decay modes, including baryonic decay modes. Baryonic $b\to s\ell^+
 \ell^-$ decays are sensitive to all possible Dirac structures in the effective Hamiltonian, and the possible polarization of the initial baryon, as well as parity-violating weak decays of final-state baryon, enable new types of angular observables \cite{Mannel:1997xy,Hiller:2001zj,Boer:2014kda,Meinel:2016grj,Blake:2017une,Das:2018sms,Yan:2019tgn,Blake:2019guk}. In this sector, the focus has been primarily on the bottom baryon with the largest production fraction, the $\Lambda_b$. The $\Lambda_b \to \Lambda \ell^+ \ell^- $ local form factors have been studied with a variety of theoretical tools in Refs.~\cite{Hussain:1990uu,Hussain:1992rb,Cheng:1994kp,Cheng:1995fe,Huang:1998ek,Mohanta:1999id,He:2006ud,Wang:2008sm,Wang:2009hra,Aliev:2010uy,Wang:2011uv,Feldmann:2011xf,Mott:2011cx,Mannel:2011xg,Gutsche:2013pp,Wang:2015ndk,Mott:2015zma,Faustov:2017wbh,Liu:2015qfa,Liu:2019igt,Yang:2025yaw,Lu:2025gjt,Mahmoudi:2026aul}, and with lattice QCD in Refs.~\cite{Detmold:2012vy,Detmold:2016pkz}. Experimental measurements of the branching fraction and angular observables of the $\Lambda_b \to \Lambda(\to p \pi) \mu^+ \mu^- $ transition \cite{LHCb:2013uqx, LHCb:2015tgy, LHCb:2018jna} were used in combination with a lattice QCD calculation of the local form factors \cite{Detmold:2016pkz} to constrain Wilson coefficients in Refs.~\cite{Meinel:2016grj,Blake:2019guk}. Next-generation lattice calculations of the $\Lambda_b \to \Lambda$ form factors are now in progress \cite{Meinel:2023wyg}.
 
 With analyses of the $\Lambda_b$ transitions reaching a more advanced stage, it is sensible to extend this baryonic perspective to another set of $b \to s \ell^+ \ell^-$ observables from the $SU(3)$ partner processes $\Xi_b \to \Xi \ell^+ \ell^-$. Although the $\Xi_b^{0,-}$ have relatively lower production fractions, there is already some experimental interest in rare decays of the $\Xi_b^-$. An LHCb analysis of the $\Xi^-_b \to \Xi^- \mu^+ \mu^-$ mode with the combined Run 1 and 2 datasets is in progress, motivated by the expectation of low mesonic backgrounds from a distinctive decay topology and good efficiency reconstructing the charged particle decay chain of the $\Xi_b^-$ \cite{Nicolini:2024opv}. Blinded measurements of the ratio $r_{BR} = \mathcal{B}(\Xi^-_b \to \Xi^- \mu^+ \mu^-)/\mathcal{B}(\Xi^-_b \to \Xi^- J/\psi(\to \mu^+ \mu^-))$ are discussed in Ref.~\cite{Nicolini:2024opv}. This promising experimental outlook motivates new theoretical work.

The individual form factors that describe the $\Xi_b\to \Xi \ell^+ \ell^-$ transitions were first estimated in the large recoil (low $q^2$) limit with light-cone sum rules in Ref.~\cite{Azizi:2011mw} (see also Refs.~\cite{Azizi:2016dcj} and \cite{Nayek:2020hna} for phenomenological applications). Furthermore, enabled by the LHCb measurements of $\mathrm{d}\mathcal{B}(\Lambda_b \to \Lambda \mu^+ \mu^- )/\mathrm{d}q^2$ \cite{LHCb:2013uqx,LHCb:2015tgy}, the values of $\mathrm{d}\mathcal{B}(\Xi^{-}_b \to \Xi^{-} \mu^+ \mu^-)/\mathrm{d}q^2$ (and other related decays) were estimated by flavor $SU(3)$ symmetry in Ref.~\cite{Wang:2021uzi}. Recently, the form factors were calculated again at low $q^2$ within the ``pQCD framework'' in Ref.~\cite{Rui:2025ajk}, leading to new Standard-Model predictions of the $\Xi_b\to \Xi \ell^+ \ell^-$ differential branching fraction and angular observables. The predictions at high $q^2$ depend on extrapolations of the form factors and have large uncertainty. A lattice QCD calculation of the form factors will be most precise at high $q^2$ and is crucial for obtaining reliable Standard-Model predictions for the $\Xi_b \to \Xi \ell^+ \ell^-$ differential branching fraction and angular observables.

It is also interesting to study the radiative decays $\Xi_b \to \Xi \gamma$, whose rates depend on the tensor form factors at $q^2=0$.
Like its better studied counterpart $\Lambda_b \to \Lambda \gamma$, which was first observed by LHCb in 2019~\cite{LHCb:2019wwi}, these processes are sensitive to the helicity structure of the $b \to s \gamma $ transition~\cite{Mannel:1997xy,Hiller:2001zj}, can be used to search for possible new sources of $CP$ violation~\cite{Hiller:2001zj}, and constrain the Wilson coefficients $C_7$ and $C_7^\prime$. A 2021 LHCb search \cite{LHCb:2021hfz} set an upper limit for the $\Xi_b \to \Xi \gamma$ branching fraction. The Standard-Model predictions of Refs.~\cite{Wang:2020wxn, Olamaei:2021eyo, Geng:2022xpn, Davydov:2022glx,Aliev:2023mdf} are below this limit, while the prediction of Ref.~\cite{Liu:2011ema} is above. A lattice calculation of the relevant $\Xi_b\to \Xi$ tensor form factors can clarify this situation.

In this work, we present the first lattice QCD calculation of the vector, axial-vector and tensor form factors governing the $\Xi_b \to \Xi\ell^+ \ell^-$ and $\Xi_b \to \Xi \gamma$ transitions, using domain-wall fermions for the light and strange quarks and an anisotropic clover action for the bottom quark. On four ensembles, with lattice spacings ranging from $0.073$ to $0.111$ \rm{fm} and pion masses ranging from $0.232$ to $0.431$ \rm{GeV}, we compute three-point functions for 15 unique source-sink separations to reliably extract the ground-state form factors. We then perform chiral-continuum-kinematic extrapolations using modified $z$ expansions, with constraints from both dispersive bounds and form factor asymptotics to achieve controlled uncertainties in the full semileptonic kinematic region. Using our form factor results, we obtain Standard-Model predictions for the $\Xi_b^- \to \Xi^- \gamma$ and $\Xi_b^- \to \Xi^- \mu^+\mu^-$ branching fractions and two angular observables.

\FloatBarrier
\section{Form-Factor Definitions}
\FloatBarrier

As in our recent calculation Ref.~\cite{Farrell:2025gis}, and as in the previous calculations of Refs.~\cite{Detmold:2015aaa, Detmold:2016pkz, Meinel:2016dqj, Meinel:2017ggx}, we define the form factors in the helicity-based scheme first described in Ref.~\cite{Feldmann:2011xf}. In this scheme, the hadronic matrix elements are written as
\begin{align}
 \nonumber \langle \Xi(p^\prime,s^\prime) | \overline{s} \,\gamma^\mu\, b | \Xi_b(p,s) \rangle = & \:
 \overline{u}_\Xi(p^\prime,s^\prime) \bigg[ f_0(q^2)\: (m_{\Xi_b}-m_\Xi)\frac{q^\mu}{q^2} \\
 & \phantom{\overline{u}_\Xi \bigg[}+ f_+(q^2) \frac{m_{\Xi_b}+m_\Xi}{s_+}\left( p^\mu + p^{\prime \mu} - (m_{\Xi_b}^2-m_\Xi^2)\frac{q^\mu}{q^2}  \right) \\
 \nonumber & \phantom{\overline{u}_\Xi \bigg[}+ f_\perp(q^2) \left(\gamma^\mu - \frac{2m_\Xi}{s_+} p^\mu - \frac{2 m_{\Xi_b}}{s_+} p^{\prime \mu} \right) \bigg] u_{\Xi_b}(p,s),
\end{align}

\begin{align}
 \nonumber \langle \Xi(p^\prime,s^\prime) | \overline{s} \,\gamma^\mu\gamma_5\, b | \Xi_b(p,s) \rangle =&
 -\overline{u}_\Xi(p^\prime,s^\prime) \:\gamma_5 \bigg[ g_0(q^2)\: (m_{\Xi_b}+m_\Xi)\frac{q^\mu}{q^2} \\
 & \phantom{\overline{u}_\Xi \bigg[}+ g_+(q^2)\frac{m_{\Xi_b}-m_\Xi}{s_-}\left( p^\mu + p^{\prime \mu} - (m_{\Xi_b}^2-m_\Xi^2)\frac{q^\mu}{q^2}  \right) \\
 \nonumber & \phantom{\overline{u}_\Xi \bigg[}+ g_\perp(q^2) \left(\gamma^\mu + \frac{2m_\Xi}{s_-} p^\mu - \frac{2 m_{\Xi_b}}{s_-} p^{\prime \mu} \right) \bigg]  u_{\Xi_b}(p,s),
\end{align}

\begin{align}
  \nonumber \langle \Xi(p^\prime,s^\prime) | \overline{s} \,i\sigma^{\mu\nu} q_\nu \, b | \Xi_b(p,s) \rangle =&
 - \overline{u}_\Xi(p^\prime,s^\prime) \bigg[  h_+(q^2) \frac{q^2}{s_+} \left( p^\mu + p^{\prime \mu} - (m_{\Xi_b}^2-m_{\Xi}^2)\frac{q^\mu}{q^2} \right) \\
 & \phantom{\overline{u}_\Xi \bigg[} + h_\perp(q^2)\, (m_{\Xi_b}+m_\Xi) \left( \gamma^\mu -  \frac{2  m_\Xi}{s_+} \, p^\mu - \frac{2m_{\Xi_b}}{s_+} \, p^{\prime \mu}   \right) \bigg] u_{\Xi_b}(p,s), 
\end{align}

\begin{align}
  \nonumber \langle \Xi(p^\prime,s^\prime)| \overline{s} \, i\sigma^{\mu\nu}q_\nu \gamma_5  \, b|\Xi_b(p,s)\rangle =&
 -\overline{u}_{\Xi}(p^\prime,s^\prime) \, \gamma_5 \bigg[   \tilde{h}_+(q^2) \, \frac{q^2}{s_-} \left( p^\mu + p^{\prime \mu} -  (m_{\Xi_b}^2-m_{\Xi}^2) \frac{q^\mu}{q^2} \right) \\
 & \phantom{\overline{u}_\Xi \bigg[}  + \tilde{h}_\perp(q^2)\,  (m_{\Xi_b}-m_\Xi) \left( \gamma^\mu +  \frac{2 m_\Xi}{s_-} \, p^\mu - \frac{2 m_{\Xi_b}}{s_-} \, p^{\prime \mu}  \right) \bigg]  u_{\Xi_b}(p,s),
\end{align}
with $q=p-p'$, $s_\pm =(m_{\Xi_b} \pm m_\Xi)^2 -q^2$, and $\sigma^{\mu\nu}=\frac{i}{2}(\gamma^\mu\gamma^\nu-\gamma^\nu\gamma^\mu)$. The helicity form factors satisfy the endpoint relations 
\begin{eqnarray}
 f_0(0) &=& f_+(0), \label{eq:endptconst1}  \\
 g_0(0) &=& g_+(0), \label{eq:endptconst2} \\
 h_\perp(0) &=& \tilde{h}_\perp(0) \label{eq:endptconst3} \\
 g_\perp(q^2_{\rm max}) &=& g_+(q^2_{\rm max}), \label{eq:endptconst4} \\
 \tilde{h}_\perp(q^2_{\rm max}) &=& \tilde{h}_+(q^2_{\rm max}), \label{eq:endptconst5} 
\end{eqnarray}
with $q^2_{\rm{max}}=(m_{\Xi_b}-m_{\Xi})^2$. 

\FloatBarrier
\section{Lattice Calculation and Correlation-Function Fits}
\FloatBarrier

\begin{table}
 \begin{tabular}{lccccccccccccc}
\hline\hline \\ [-2.8ex]
Label & $N_s^3\times N_t \times N_5$ & $\beta$  & $a$ [fm] &  $am_{u,d}$ &  $m_\pi$ [GeV] & $m_\pi L$ & $am_{s}^{(\mathrm{sea})}$ 
& $am_{s}^{(\mathrm{val})}$ & $\wm a m_Q^{(b)}$ & $\nu^{(b)}$ & $c_{E,B}^{(b)}$  & $N_{\rm ex}$ & $N_{\rm sl}$ \\
\hline
C01  & $24^3\times64\times16$ & $2.13$   & $0.1106(3)$ & $0.01\nb$  & $0.4312(13)$ & 5.813(12) & $0.04$      & $0.0323$  & $7.3258$ & $3.1918$ & $4.9625$  & 283 & 2264  \\
C005 & $24^3\times64\times16$ & $2.13$   & $0.1106(3)$ & $0.005$    & $0.3400(11)$ & 4.570(14) & $0.04$      & $0.0323$  & $7.3258$ & $3.1918$ & $4.9625$  & 311 & 2488  \\
F004 & $32^3\times64\times16$ & $2.25$   & $0.0828(3)$ & $0.004$    & $0.3030(12)$ & 4.061(13) & $0.03$      & $0.0248$  & $3.2823$   & $2.0600$ & $2.7960$  & 251 & 2008  \\
F1M  & $48^3\times96\times12$ & $2.31$   & $0.0728(3)$ & $0.002144$ & $0.2320(10)$ & 4.116(8) & $0.02144$   & $0.02217$ & $2.3867$  & $1.8323$ & $2.4262$ & 113 & 1808 \\
\hline\hline
\end{tabular}
\caption{Lattice and action parameters for each of the ensembles. The ensemble generation and lattice spacings are discussed in Refs.~\cite{RBC:2010qam,RBC:2014ntl,Boyle:2018knm}. The light valence quarks are implemented with the same action as the sea quarks, with masses set equal to the sea quark values. The strange valence quarks are also implemented with the same action, but with masses tuned to the physical point. The F1M ensemble uses a M\"obius domain-wall action~\cite{Boyle:2018knm}, while C01, C005, and F004 use a Shamir domain-wall action \cite{RBC:2010qam,RBC:2014ntl}. The bottom quark mass $a m_Q^{(b)}$, the anisotropy parameter $\nu^{(b)}$, and the chromoelectric/chromomagnetic clover coefficients $c_{E}^{(b)}=c_{B}^{(b)}$ values and tunings are discussed in Ref.~\cite{Meinel:2023wyg}. $N_{\rm ex}$ and $N_{\rm sl}$ are the number of exact and sloppy (computed with reduced conjugate-gradient iteration counts) samples used in the all-mode-averaging procedure \cite{Blum:2012uh, Shintani:2014vja}.
}
\label{tab:lattice_params}
\end{table}

This calculation was performed with the same set of ensembles and actions as in our recent work~\cite{Farrell:2025gis}. For the light quarks, we use the 2+1 flavor domain-wall fermion, Iwasaki gauge action ensembles generated by the RBC and UKQCD collaborations~\cite{RBC:2010qam,RBC:2014ntl,Boyle:2018knm}, while for the heavy quark we again use an anisotropic clover action analogous to Ref.~\cite{RBC:2012pds}. The parameters $a m_Q^{(b)}$, $\nu^{(b)}$, and $c_{E}^{(b)}=c_{B}^{(b)}$ of the latter are tuned nonperturbatively to give the experimental values of the $B_s$ meson rest mass, kinetic mass, and hyperfine splitting. The tuning procedure is detailed in Ref~\cite{Meinel:2023wyg} (for the bare parameters we use the notation of Ref.~\cite{Brown:2014ena}, whereas Ref.~\cite{RBC:2012pds} uses $m_0=m_Q$, $\zeta=\nu$, $c_P=c_E=c_B$). The major lattice and action parameters for each of the four ensembles are listed in Table~\ref{tab:lattice_params}. 


For the $\Xi_b$ and $\Xi$ baryons, we use the interpolating fields
\begin{align}
    \Xi_{b{\alpha}} = \epsilon_{abc} (C\gamma_5)_{\beta\gamma} d^a_\beta s^b_\gamma b^c_\alpha, \hspace{1em}
    \Xi_{\alpha} = \epsilon_{abc} (C\gamma_5)_{\beta\gamma} d^a_\beta s^b_\gamma s^c_\alpha. 
\end{align}
The smearing parameters for the quark fields are listed in Table~\ref{tab:smearingparams}. We compute the forward and backward two-point correlation functions
\begin{equation}
\begin{aligned}
        C^{(2,\Xi,\text{fw})}_{\delta \alpha}(\textbf{p}',t) &= \sum_{\textbf{y}} e^{-i\textbf{p}' \cdot (\textbf{y}-\textbf{x})} \big<\Xi_{\delta}(x_0+t, \textbf{y}) \: \overline{\Xi}_\alpha(x_0,\textbf{x}) \big>, \\
        C^{(2,\Xi,\text{bw})}_{\delta \alpha}(\textbf{p}',t) &= \sum_{\textbf{y}} e^{-i\textbf{p}' \cdot (\textbf{x}-\textbf{y})} \big<\Xi_{\delta}(x_0, \textbf{x}) \: \overline{\Xi}_\alpha(x_0-t,\textbf{y}) \big>,\\
        C^{(2,\Xi_b,\text{fw})}_{\delta \alpha}(t) &= \sum_{\textbf{y}} \big<\Xi_{b\delta}(x_0+t, \textbf{y}) \: \overline{\Xi}_{b\alpha}(x_0,\textbf{x}) \big>, \\
        C^{(2,\Xi_b,\text{bw})}_{\delta \alpha}(t) &= \sum_{\textbf{y}} \big<\Xi_{b\delta}(x_0, \textbf{x}) \: \overline{\Xi}_{b\alpha}(x_0-t,\textbf{y}) \big>,
    \end{aligned}
\end{equation}
and the forward and backward three-point correlation functions
\begin{equation}
    \begin{aligned}
        C^{(3,\text{fw})}_{\delta \alpha}(\Gamma,\mathbf{p}',t,t') = \sum_{\mathbf{y},\mathbf{z}} e^{-i\mathbf{p}' \cdot (\mathbf{x}-\mathbf{y})} \Big<\Xi_{\delta}(x_0, \mathbf{x}) \:\: J_\Gamma(x_0 -t + t', \mathbf{y}) \:\:  \overline{\Xi}_{b\alpha}(x_0-t,\mathbf{z}) \Big>, \\
        C^{(3,\text{bw})}_{\delta \alpha}(\Gamma,\mathbf{p}',t,t-t') = \sum_{\mathbf{y},\mathbf{z}} e^{-i\mathbf{p}' \cdot (\mathbf{y}-\mathbf{x})} \Big<\Xi_{b\delta}(x_0 + t, \mathbf{z}) \:\: J^{\dagger}_\Gamma(x_0 + t', \mathbf{y}) \:\:  \overline{\Xi}_\alpha(x_0,\mathbf{x}) \Big>,
    \end{aligned}
    \label{eq:threepoint}
\end{equation}
where $J_\Gamma$ is discussed in Eq.~(\ref{eq:current}) below. We take our initial state $\Xi_b$ to have zero spatial momentum and use sequential propagators for the bottom quark in the three-point correlation functions. The correlation functions are computed with light and strange quark propagators with source position $(x_0,\mathbf{x})$, which are reused from Ref.~\cite{Meinel:2020owd} for C01, C005, F004 and from Ref.~\cite{Meinel:2023wyg} for F1M.

\begin{table}
	\begin{tabular}{lccccccccc} \hline \hline 
		Ensemble   & \multicolumn{2}{c}{Up and down quarks} & \hspace{2ex} & \multicolumn{2}{c}{Strange quarks} & \hspace{2ex} & \multicolumn{2}{c}{Bottom quarks} \\
               & $N_\textrm{Gauss}$ & $\sigma_\textrm{Gauss}/a$ & \hspace{2ex}  &  $N_\textrm{Gauss}$ & $\sigma_\textrm{Gauss}/a$ &  \hspace{2ex} & $N_\textrm{Gauss}$ & $\sigma_\textrm{Gauss}/a$  \\ \hline
    C005, C01  & $30$ & $4.350$ & \hspace{2ex}  & $30$ & $4.350$ & \hspace{2ex}  & $10$ & $2.0$ \\
    F004 & $60$ & $5.728$           & \hspace{2ex}  & $60$ & $5.728$ & \hspace{2ex}  & $16$ & $2.667$ \\ 
    F1M & $130$ & $8.9$           & \hspace{2ex}  & $70$ & $6.6$ & \hspace{2ex}  & $20$ & $3.0$ \\ \hline \hline
	\end{tabular}
	\caption{\label{tab:smearingparams}Parameters for the smearing of the quark fields in the baryon interpolating fields. The Gaussian smearing is defined as in Eq.~(8) of Ref.~\cite{Leskovec:2019ioa}. For the up, down, and strange quarks, we used APE-smeared gauge links \cite{Bonnet:2000dc} with $N_{\rm APE}=25$ and $\alpha_{\rm{APE}}=2.5$ in the Gaussian smearing kernel. For the bottom quarks, we use Stout-smeared \cite{Morningstar:2003gk} gauge links with $N_{\rm Stout}=10$ and $\rho_{\rm Stout}=0.08$ in the Gaussian smearing.}
\end{table}


The $b \to s$ currents $J_\Gamma$ are structured as
\begin{equation}
    J_\Gamma = \rho_\Gamma \sqrt{Z^{ll}_{V}Z^{bb}_{V}}\bigg[\bar{s}\Gamma b + \mathcal{O}(a) \text{-improvement terms} \bigg] \label{eq:current},
\end{equation}
with the complete forms of the vector and axial vector currents with the $\mathcal{O}(a)$-improvement terms are given in Eqs.~(18)-(21) of  Ref.~\cite{Detmold:2015aaa}, while the tensor current is given explicitly in Eq.~(24) of Ref.~\cite{Detmold:2016pkz}. We again use the mostly nonperturbative renormalization scheme~\cite{Hashimoto:1999yp,El-Khadra:2001wco}, in which most of the renormalization is captured in the renormalization factors of the temporal parts of the light-to-light vector current ($Z^{ll}_{V}$) and bottom-to-bottom vector current ($Z^{bb}_{V}$), which are determined nonperturbatively using charge conservation. The values of these factors for each ensemble are given in Table~\ref{tab:ZV}. 

Table~\ref{tab:Pmatchingfactors} lists both the residual matching factors $\rho_\Gamma$ and the $\mathcal{O}(a)$-improvement coefficients for the vector and axial vector currents. For the tensor currents, one-loop results for these coefficients were not available, and we used the mean-field improved tree-level values as in Ref.~\cite{Detmold:2016pkz}, with the values of $d_1$ also given in Table~\ref{tab:Pmatchingfactors}. This contributes an additional source of systematic uncertainty, which is discussed in detail in Section~\ref{sec:extrap}.

\begin{table}
\begin{center}
\small
\begin{tabular}{lllllllll}
\hline\hline \\[-2.5ex]
Ensemble          & \hspace{2ex} & $Z_V^{(ll)}$  & \hspace{2ex} & $Z_V^{(bb)}$     \\
\hline
  C005, C01     &&  $0.71273(26)$ \cite{RBC:2014ntl}        &&  $9.0631(84)$ \cite{Meinel:2023wyg}      \\[0.2ex]
  F004          &&  $0.7440(18)$  \cite{RBC:2014ntl}        &&  $4.7449(21)$ \cite{Meinel:2023wyg}      \\[0.2ex]
  F1M           &&  $0.7639(42)$  \cite{Marshall:2024pfg}   &&  $3.7777(23)$ \cite{Meinel:2023wyg}     \\[0.2ex]
\hline\hline
\end{tabular}
\caption{\label{tab:ZV} Nonperturbative results for the renormalization factors of the temporal components of the flavor-conserving vector currents.}
\end{center}
\end{table}

\begin{table}
\begin{center}
\small
\begin{tabular}{clllllll}
\hline\hline
Parameter         & \hspace{2ex}            & \hspace{1ex}  C005, C01     & \hspace{2ex} & \hspace{1ex} F004   & \hspace{2ex} &  \hspace{1ex} F1M   \\
\hline
 $\rho_{V^0}=\rho_{A^0}$     & & $\wm1.027(10)$       & &  $\wm 1.0166(51)$      & &  $\wm 1.0112(74)$     \\[0.5ex] 
 $\rho_{V^j}=\rho_{A^j}$     & & $\wm0.9972(11)$      & &  $\wm 0.9940(18)$      & &  $\wm 0.9922(25)$     \\[0.5ex] 
 $c_{V^0}^R=c_{A^0}^R$       & & $\wm0.0558(73)$      & &  $\wm 0.0547(57)$      & &  $\wm 0.0541(58)$     \\[0.5ex] 
 $c_{V^0}^L=c_{A^0}^L$       & & $-0.0099(39)$        & &  $-0.0095(29)$         & &  $-0.0093(29)$        \\[0.5ex] 
 $c_{V^j}^R=c_{A^j}^R$       & & $\wm0.0485(45)$      & &  $\wm 0.0480(37)$      & &  $\wm 0.0477(37)$     \\[0.5ex] 
 $c_{V^j}^L=c_{A^j}^L$       & & $-0.0033(13)$        & &  $-0.00200(61)$        & &  $-0.00129(93)$       \\[0.5ex] 
 $d_{V^j}^R=-d_{A^j}^R$      & & $-0.00079(31)$       & &  $-0.00120(37)$        & &  $-0.00142(43)$       \\[0.5ex] 
 $d_{V^j}^L=-d_{A^j}^L$      & & $\wm0.00180(70)$     & &  $\wm 0.00047(14)$     & &  $-0.00026(74)$       \\[0.5ex] 
 $d_1$                       & & $\wm 0.07403$     & &  $\wm0.07182$       & &  $\wm0.06797$     \\[0.5ex]
\hline\hline
\end{tabular}\vspace{-2ex}
\end{center}
\caption{\label{tab:Pmatchingfactors} Perturbative renormalization and $\mathcal{O}(a)$-improvement coefficients for the $b\to s$ currents used in Eq.~(\ref{eq:current}). For the C005/C01 and F004 ensembles, they were computed by C.~Lehner at one loop in mean-field-improved lattice perturbation theory \cite{Lehner:2012bt}. For these two ensembles, the uncertainties given in Table \ref{tab:Pmatchingfactors} correspond to the change in the central value when changing the strong coupling from a mean-field lattice $\overline{\rm MS}$ coupling $\alpha_{s,{\rm lat}}^{\overline{\rm MS}}(a^{-1})$ to the continuum $\overline{\rm MS}$ coupling $\alpha_{s,{\rm ctm}}^{\overline{\rm MS}}(a^{-1})$ \cite{Flynn:2023nhi}. For the newer F1M ensemble, perturbative results were not available, and therefore we estimated the coefficients through an extrapolation of the values from the C005/C01 and F004 ensembles, done linearly in the lattice spacing. In this case, the uncertainties given in the table are the sum (in quadrature) of the F004 uncertainties and the shifts in the central values between F004 and F1M. Doing the extrapolation of the coefficients quadratically instead of linearly in the lattice spacing changes the values much less than the estimated uncertainties.} 
\end{table}
\begin{figure}
    \vspace{-6ex}
    \centering
    \includegraphics[scale=.75]{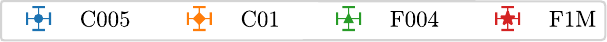}
    \includegraphics[scale=.75]{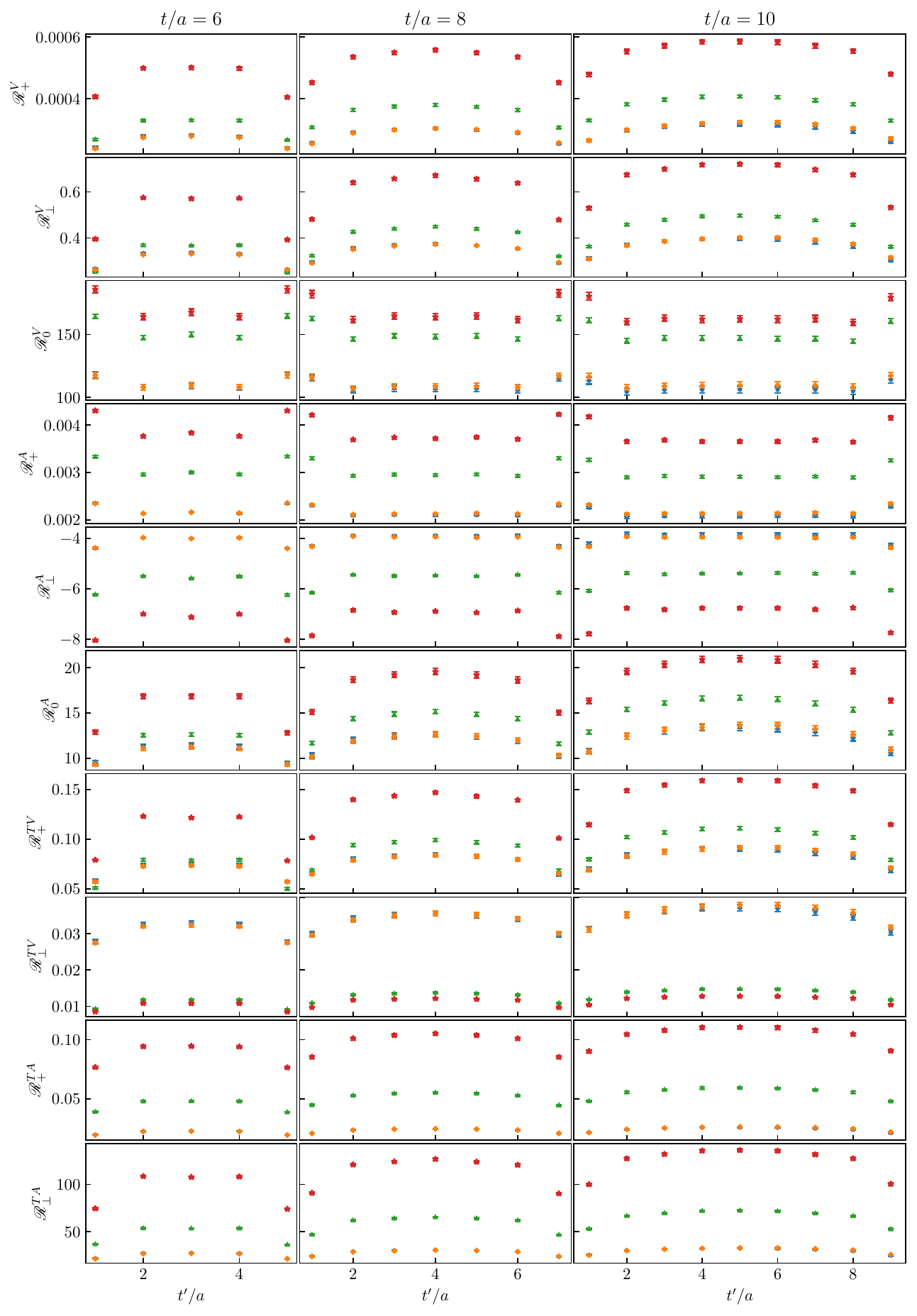}
    \caption{ The $t'$ dependence of the different ratios with vector, axial vector and tensor current insertions for three different values of the source-sink separation, $t$. The data from the C005, C01, and F004 ensembles are shown at $|\mathbf{p}^\prime|^2=1 \cdot (\tfrac{2\pi}{L})^2$, and the data from the F1M ensemble are shown at $|\mathbf{p}^\prime|^2=2\cdot (\tfrac{2\pi}{L})^2$. The plots are in units of $\rm{GeV}^{-2}$ for the dimensionful ratios $(\mathscr{R}^V_\perp, \mathscr{R}^V_0, \mathscr{R}^A_\perp, \mathscr{R}^A_0,\mathscr{R}^{TV}_+,\mathscr{R}^{TA}_+)$ and of $\rm{GeV}^{-4}$ for $( \mathscr{R}^{TV}_\perp,\mathscr{R}^{TA}_\perp)$; the uncertainty from the lattice spacing is not shown.}
    \label{fig:tpdep}
\end{figure}

As in Refs.~\cite{Detmold:2015aaa, Detmold:2016pkz, Meinel:2016dqj, Meinel:2017ggx, Farrell:2025gis}, we compute the helicity form factors from ratios of three-point and two-point correlation functions that are designed to cancel the overlap factors and time dependence of the ground-state contribution. For example, the $h_+$ form factor is obtained from the ratio 
\begin{eqnarray}
\mathscr{R}_{+}^{TV}(\mathbf{p}^\prime,t,t^\prime) &=& \frac{ r_\mu[(1,\mathbf{0})] \: r_\nu[(1,\mathbf{0})] \:
\mathrm{Tr}\Big[   C^{(3,{\rm fw})}(\mathbf{p'},\:i\sigma^{\mu\rho}q_\rho, t, t') \:    C^{(3,{\rm bw})}(\mathbf{p'},\:i\sigma^{\nu\lambda}q_\lambda, t, t-t')  \Big] }
{\mathrm{Tr}\Big[C^{(2,\Xi,{\rm av})}(\mathbf{p'}, t)\Big] \mathrm{Tr}\Big[C^{(2,\Xi_b,{\rm av})}(t)\Big] },  \label{eq:ratio1}
\end{eqnarray}
where we use the averages of the forward and backward two-point functions, and where $r_\mu[(1,\mathbf{0})]$ is a polarization vector that projects to the desired helicity, defined as
\begin{align}
    r[n]= n- \frac{(q \cdot n)}{q^2}q.
\end{align}
The dependence on the current insertion time $t'$ of $\mathscr{R}_{+}^{TV}$ and the other ratios is shown in Fig.~\ref{fig:tpdep}. From these ratios we construct new quantities such as, in the case of the form factor $h_+$,
\begin{eqnarray}
R_{h_+}(|\mathbf{p}^\prime|, t)     &=& \frac{2}{E_\Xi-m_\Xi} \sqrt{ \frac{ E_\Xi}{ (E_\Xi+m_\Xi)} \mathscr{R}_{+}^{TV}(|\mathbf{p}^\prime|, t, t/2)}, \label{eq:Rhplus}
\end{eqnarray}
which, because the excited-state contamination is suppressed at large Euclidean time, asymptotically approach the value of the form factor. We average over the directions of $\mathbf{p}^\prime$, while the $t'$ dependence is removed by evaluating the ratio at $t'=t/2$ (or averaging over the two mid-points for odd $t/a$). The values of $m_{\Xi_b}$ and $m_{\Xi}$ on each ensemble are given in Table~\ref{tab:hadronmasses}. The quantities $R_{f_+}$, $R_{f_\perp}$, $R_{f_0}$, $R_{g_+}$, $R_{g_\perp}$, and $R_{g_0}$ are given explicitly in Eqs.~(46)-(60) of Ref.~\cite{Detmold:2015aaa}, while $R_{h_+}$, $R_{h_\perp}$, $R_{\tilde{h}_+}$, and $R_{\tilde{h}_\perp}$ are given in Eqs.~(27)-(30) of Ref.~\cite{Detmold:2016pkz}. These were computed at source-sink separations of $t/a=4,5,...,15$ on the C01 and C005 ensembles, $t/a=5,6,...,15$ on F004 and $t/a=6,7,...,20$ on F1M. We computed the ratios for values of $\Xi$ spatial momenta squared of $\{1,2,3,4,5,6,8,9,10\} \cdot (\tfrac{2\pi}{L})^2$ on C005, C01, F004 and $\{1,2,3,4,5,6,8,9,10,11,12,13,14,16\} \cdot (\tfrac{2\pi}{L})^2$ on F1M.
\begin{table}
\begin{tabular}{ccccccccccc}
\hline\hline
Ensemble  & \hspace{1ex} & $a m_{\Xi_b}$ & \hspace{1ex} & $am_{\Xi}$  & \hspace{1ex} & $am_{B}$ & \hspace{1ex} & $am_{K}$     \\
\hline
C01     &&  $3.2956(33)$  &&  $0.8078(23)$  &&  $2.9792(19)$  &&  $0.32457(45)$  \\
C005    &&  $3.2718(38)$  &&  $0.7821(22)$  &&  $2.9729(20)$   &&  $0.30706(54)$  \\
F004    &&  $2.4447(42)$  &&  $0.5832(14)$  &&  $2.2248(10)$   &&  $0.22492(58)$  \\ 
F1M     &&  $2.1485(27)$  &&  $0.5011(12)$  &&  $1.9545(13)$   &&  $0.19073(19)$  \\ 
\hline\hline
\end{tabular}
\caption{\label{tab:hadronmasses}Hadron masses in lattice units. The $B$ and $K$ masses are used in the chiral-continuum-kinematic extrapolations of the form factors.}
\end{table}

To extrapolate the functions $R_f$ to infinite source-sink separation, we use the same fit functions (for each form factor $f$, each value of $n=|\mathbf{p}^\prime|^2/(2\pi/L)^2$, and each ensemble $i$) as in our previous calculations \cite{Farrell:2025gis, Detmold:2015aaa, Detmold:2016pkz, Meinel:2016dqj, Meinel:2017ggx}:
\begin{align}
    R_{f,i,n}(t) = f_{i,n} + A_{f,i,n} \: e^{-\delta E_{f,i,n}\:t}, \hspace{2ex} \delta E_{f,i,n}=\delta E_{\rm min} + e^{\,l_{f,i,n}}\:\:{\rm GeV},
    \label{eq:ratiofitfunc}
\end{align}
parameterized by $f_{i,n}$, $A_{f,i,n}$, and $l_{f,i,n}$. The dominant excited-state contamination is captured in the exponential term, while $f_{i,n}$ is the ground-state value of the form factor. As we did in Ref.~\cite{Farrell:2025gis}, we set the minimum energy gap $\delta E_{\rm min}$ to 100 MeV. To ensure the energy gap parameters $l_{f,i,n}$ do not widely vary between the different ensembles with the same value of $L$, they are again constrained by priors, as in Eq.~(70) of Ref.~\cite{Detmold:2015aaa}. For some data points at higher momenta where the quantities $R_{f,i,n}$ had larger error bars and did not display any curvature, the exponential term was omitted from the fit function. 

As previously in Ref.~\cite{Farrell:2025gis, Detmold:2015aaa, Detmold:2016pkz, Meinel:2016dqj, Meinel:2017ggx}, the data for the three vector form factors ($f=f_+,f_\perp, f_0$), the three axial-vector form factors ($f=g_+,g_\perp, g_0$), as well as ($f=h_+,h_\perp$) and ($f=\tilde{h}_+,\tilde{h}_\perp$), are fitted simultaneously for a given momentum index $n$. These fits again also include data in the Weinberg basis to stabilize the extrapolations.

\begin{figure}
    \centering
    \includegraphics[width=\linewidth]{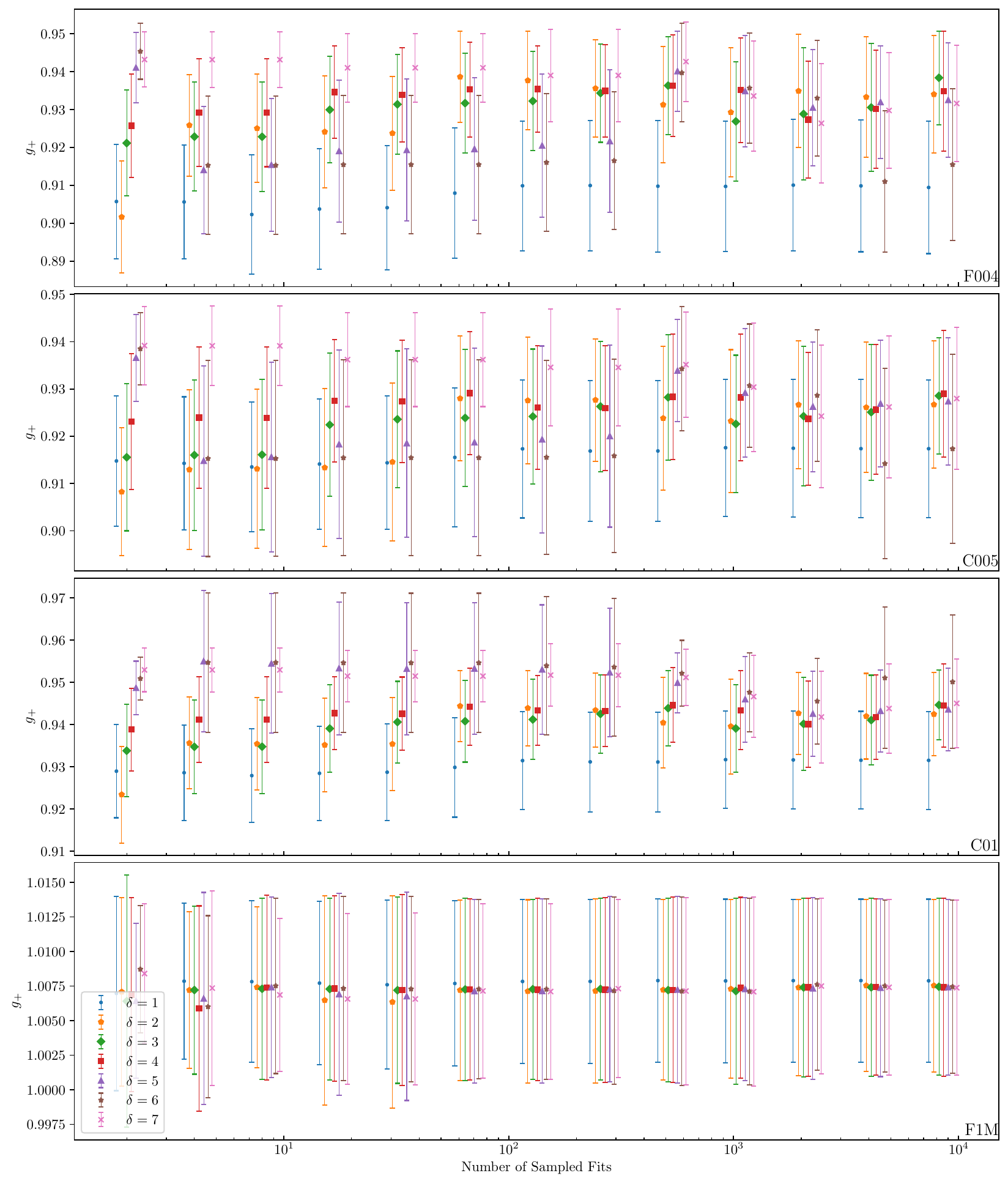}
    \caption{Evolution of the AIC average and uncertainty of the form factor $g_+$ at $|\mathbf{p}^\prime|^2=1\cdot (\tfrac{2\pi}{L})^2$ as a function of the number of sample fits, for widths of the $t_{\rm min}$ random distributions equal to $\delta=1,...,7$ (in lattice units). As in Ref.~\cite{Farrell:2025gis}, we found that values of $\delta > 4$ could require larger numbers of sample fits for the AIC average to converge, but still tended toward consistent central values. As before, we use $\delta=4$ and $\mathcal{O}(10,000)$ sample fits to obtain the final estimates.
    }
    \label{fig:delta_dep}
\end{figure}
We estimate the systematic uncertainties from these fits with the same ``perfect model'' Akaike information criterion (AIC) Refs.~\cite{Jay:2020jkz, Neil:2023pgt} that we utilized in our recent work on $\Xi_c\to\Xi$ \cite{Farrell:2025gis}. As described in Refs.~\cite{Jay:2020jkz, Neil:2023pgt}, the AIC average for a fit parameter is a weighted average with the weights determined by an information criterion (in this case the ``perfect model'' AIC). The variance of the AIC average, computed using Eq.~(24) of Ref.~\cite{Farrell:2025gis}, includes the systematic uncertainty resulting from the choice of fit ranges.  
\begin{figure}
    \centering
    \includegraphics[width=\linewidth]{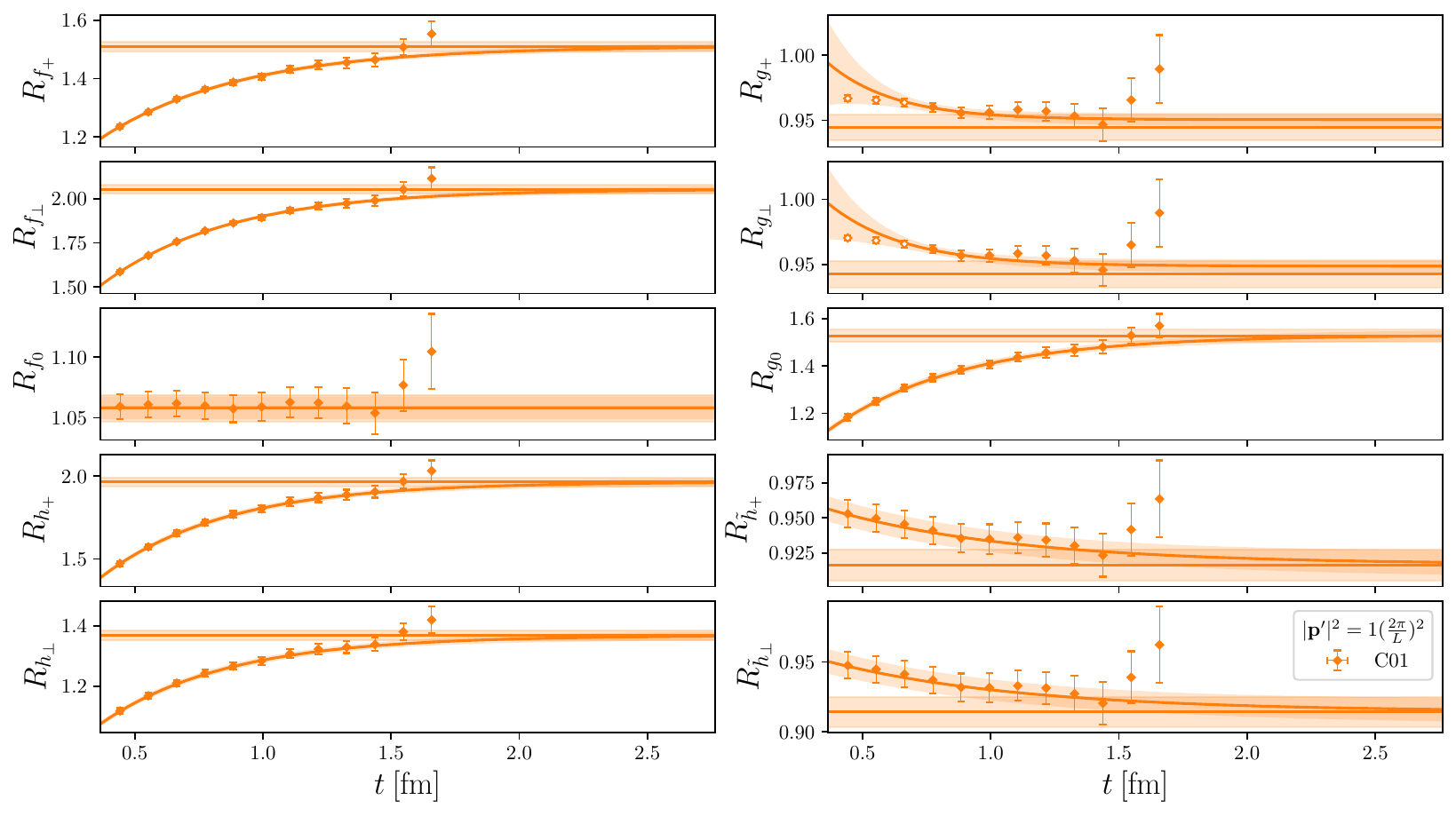}
    \caption{AIC fit results for $R_f(|\mathbf{p}^\prime|,t)$ on the $\rm{C01}$ ensemble at $|\mathbf{p}^\prime|^2=1(2\pi/L)^2$. The curves going through the data points belong to the sample fit with the highest model weight, with the bands showing only the statistical uncertainty, whereas the horizontal bands depict to the AIC average values of the extracted ground-state form factors and their total uncertainty. Data points plotted with open symbols were omitted in the highest-model-weight fit.}  \label{fig:c01_ratios}
\end{figure}
We follow a similar prescription as in Ref.~\cite{Farrell:2025gis}, and randomly sample the high-dimensional space of $t_{\rm min}$'s for each simultaneous fit. For every individual form factor, we draw each $t_{\rm min}$ from a uniform random distribution ranging from a value equal to or near the smallest value for which we have data, to a value that is larger by some chosen $\delta$ (so $\delta$ corresponds to the width of the random distribution of $t_{\rm min}$). Performing a large enough number of these fits with randomly selected $t_{\rm min}$'s allows us to estimate the AIC average. We again investigated the dependence of the AIC average estimate on the choice of $\delta$ and the number of sampled fits, which are shown for an example form factor in Fig.~\ref{fig:delta_dep}. As in Ref.~\cite{Farrell:2025gis}, the final estimates were obtained with $\delta=4$ and $\mathcal{O}(10,000)$ fits. We again found consistent AIC average values for $\delta \geq 2$ and ($\geq$ 7,000) random fits. As in Ref~\cite{Farrell:2025gis}, for some of the AIC extrapolated values at high values of $q^2$, the term that captures the systematic error in the AIC average [Eq.~(24) of Ref.~\cite{Farrell:2025gis}], was often far smaller than at low values of $q^2$. Therefore, to be more conservative, for all form factors on all ensembles, we computed the average of that term across the entire $q^2$ range and replaced any value that fell below it by the average value. 

Sample results of this fit procedure are shown in Fig.~\ref{fig:c01_ratios} for the ensemble $\rm{C01}$ at $|\mathbf{p}'|^2 = 1\cdot (\tfrac{2\pi}{L})^2$. Both the AIC average value (and uncertainty) and the results of the highest-weight fit are shown for each of the form factors, while example fits for the other ensembles are shown in Fig.~\ref{fig:other_ensemble_ratios} in the Appendix.

{Given that the shortest source-sink separations entering in the AIC averages are as low as $\approx 0.4$ fm, a possible concern is that the averaging procedure may not fully cover residual bias from the assumed excited-state form (\ref{eq:ratiofitfunc}). As a further test of this assumption, we performed two-state fits of the two-point correlation functions, where for each ensemble the values of $t_{\rm min}$ were chosen to be approximately half the shortest source-sink separation used in the three-point correlation functions. These fits were found to have good quality, as shown in Appendix~\ref{sec:effmass}. 

\FloatBarrier
\section{Chiral-continuum-kinematic extrapolations}
\label{sec:extrap}
\FloatBarrier

We use BGL-type series expansions \cite{Boyd:1994tt,Blake:2022vfl,Flynn:2023qmi} to parametrize the $q^2$ dependence of the form factors in the physical limit ($a=0$, $m_{\pi} =m_{\pi, \rm{ phys}}$). We define
\begin{equation}
z(q^2, t_0, t_+) = \frac{\sqrt{t_+-q^2}-\sqrt{t_+-t_0}}{\sqrt{t_+-q^2}+\sqrt{t_+-t_0}}.
\end{equation}
This transformation maps the complex $q^2$ plane onto the unit disk, with the parameter $t_0$ defining which particular value of $q^2$ is mapped to $z=0$; we use
\begin{equation}
t_0 = q^2_{\rm max} = (m_{\Xi_b} - m_{\Xi})^2.
\end{equation}
The parameters $t_{+}$ are set to the positions at which the two-particle branch cuts along the real $q^2$ axis begin, which is different for the different form factors. Since our calculation is performed in the limit of exact isospin symmetry, we have
\begin{align}
\nonumber t^{f_+,f_\perp,f_0,h_+,h_\perp}_+ &= (m_B + m_K)^2, \\
t^{g_+,g_\perp,g_0,\tilde{h}_+,\tilde{h}_\perp}_+ &= (m_{B^*} + m_K)^2. \label{eq:tplus}
\end{align}
We fit the lattice data with the functions
\begin{eqnarray}
\nonumber f(q^2) &=& \frac{1}{\mathcal{P}_f(q^2)\phi_f(q^2)}
\bigg[ a_0^f\bigg(1+c_0^f \frac{m_\pi^2-m_{\pi,{\rm phys}}^2}{\Lambda_\chi^2}+\tilde{c}_0^f \frac{m_\pi^3-m_{\pi,{\rm phys}}^3}{\Lambda_\chi^3}\bigg) \\
\nonumber && \hspace{15ex} +\: a_1^f\bigg(1+c_1^f\frac{m_\pi^2-m_{\pi,{\rm phys}}^2}{\Lambda_\chi^2}+\tilde{c}_1^f \frac{m_\pi^3-m_{\pi,{\rm phys}}^3}{\Lambda_\chi^3}\bigg)\:z(q^2,q^2_{\rm max},t^f_+)  + \sum_{n=2}^{N+M} a_n^f\:[z(q^2,q^2_{\rm max},t^f_+)]^n  \bigg] \\
 && \times\: \bigg[1  + b^f\, a^2|\mathbf{p^\prime}|^2 + d^f\, a^2\Lambda_{\rm had}^2
                     + \tilde{b}^f\, a^4 |\mathbf{p^\prime}|^4
                     + \hat{d}^f\, a^3 \Lambda_{\rm had}^3
                     + \tilde{d}^f a^4 \Lambda_{\rm had}^4
                     + j^f   a^4 |\mathbf{p^\prime}|^2\Lambda_{\rm had}^2 \bigg], \hspace{5ex}  \label{eq:FFfit}
\end{eqnarray}
which in the physical limit ($a=0$, $m_\pi=m_{\pi,{\rm phys}}$) simplify to the form
\begin{equation}
 f(q^2) = \frac{1}{\mathcal{P}_f(q^2)\phi_f(q^2)} \sum_{n=0}^{N+M} a_n^f\:[z(q^2,q^2_{\rm max},t^f_+)]^n.\label{eq:zphys}
\end{equation}
Here, the functions $\mathcal{P}_f(q^2)$ are the Blaschke factors
\begin{equation}
\mathcal{P}_f(q^2) = z(q^2, [m_{\rm pole}^f]^2, t_+^f),
\end{equation}
with the pole masses listed in Table~\ref{tab:polemasses}. The functions $\phi_f(q^2)$ are the \emph{outer functions}, which were constructed in Ref.~\cite{Blake:2022vfl} and are given in Appendix \ref{sec:outerfuncs}. The scale factors $\Lambda_{\rm had} = 300$ MeV and $\Lambda_\chi = 4\pi f_\pi$ (with $f_\pi =132$ MeV) allow all parameters to remain dimensionless. When doing the fit using Eq.~(\ref{eq:FFfit}), we use the lattice hadron masses from each individual data set\footnote{Except for the $B^*$, whose masses we evaluate by adding a fixed hyperfine splitting of 45 MeV to the lattice $B$ masses.}, to evaluate $z(q^2,q^2_{\rm max},t^f_+)$ and the outer functions, but we use the physical hadron masses from experiment \cite{ParticleDataGroup:2024cfk}, $m_{\Xi_b^-}=5.797$ GeV, $m_{\Xi^-}=1.3217$ GeV, $m_B=5.279$ GeV, $m_{B^*}=5.324$ GeV, $m_K=0.494$ GeV,  to evaluate the Blaschke factors.
When evaluating the physical-limit form factors using Eq.~(\ref{eq:zphys}), we use the physical hadron masses throughout.

In our fit using Eq.~(\ref{eq:FFfit}), we implement the endpoint constraints (\ref{eq:endptconst4}) and (\ref{eq:endptconst5}), which apply to the point $z=0$, by eliminating $a_0^{g_\perp}$ and $a_0^{\tilde{h}_\perp}$ through
\begin{eqnarray}
a_0^{g_\perp} &=& \frac{\phi_{g_\perp}(t_0)}{\phi_{g_+}(t_0)} a_0^{g_+}, \label{eq:a0gperp} \\
a_0^{\tilde{h}_\perp} &=& \frac{\phi_{\tilde{h}_\perp}(t_0)}{\phi_{\tilde{h}_+}(t_0)} a_0^{\tilde{h}_+}, \label{eq:a0htildeperp} 
\end{eqnarray}
where the outer functions are evaluated using the physical masses. We enforce the endpoint constraints (\ref{eq:endptconst1}), (\ref{eq:endptconst2}), (\ref{eq:endptconst3}) by adding the term
\begin{equation}
\frac{(f_+(0)-f_0(0))^2}{\sigma_{\rm EP}^2} + \frac{(g_+(0)-g_0(0))^2}{\sigma_{\rm EP}^2} + \frac{(h_\perp(0)-\tilde{h}_\perp(0))^2}{\sigma_{\rm EP}^2}
\end{equation}
to the $\chi^2$ function used in the fit, where the form factors are evaluated in the physical limit using Eq.~(\ref{eq:zphys}), and $\sigma_{\rm EP}$ was set to $5\cdot 10^{-5}$. Because the form factors at $q^2=0$ depend on all $z$-expansion coefficients, this method is more convenient than parameter elimination, due to the additional asymptotic-behavior constraints discussed in the following.

A known problem with BGL-type $z$ expansions is that the asymptotic behavior of the outer functions allows the form factors to grow for $q^2\to -\infty$ ($z\to 1$), which is inconsistent with the expectation from perturbative QCD \cite{Bourrely:2008za}. Given the outer functions from Ref.~\cite{Blake:2022vfl}, we find that the parametrization (\ref{eq:zphys}) behaves, for $q^2\to-\infty$, like a polynomial containing the powers $\{-q^2, \sqrt{-q^2}, 1, \frac{1}{\sqrt{-q^2}}, ...\}$ for the form factors $f_+$, $g_+$, $h_\perp$, and $\tilde{h}_\perp$, and like a polynomial containing $\{\sqrt{-q^2}, 1, \frac{1}{\sqrt{-q^2}}, ...\}$ for the remaining form factors. We therefore impose sum rules on the coefficients $a_n^f$ \cite{Buck:1998kp,Becher:2005bg,Flynn:2023qmi} such that the form factors must fall off at least like $1/(-q^2)$ for $q^2\to-\infty$, which matches the asymptotic behavior of the standard BCL-type $z$ expansion \cite{Bourrely:2008za} as used in Refs.~\cite{Detmold:2015aaa, Detmold:2016pkz}. The asymptotic behavior of heavy-to-light baryon form factors was studied, for example, in Refs.~\cite{Feldmann:2011xf,Wang:2011uv,Khodjamirian:2011jp,Mannel:2011xg,Zhou:2026fui}, which predict even faster fall-off at the perturbative orders considered. To implement these sum rules, we follow Ref.~\cite{Flynn:2023qmi} and solve for the coefficients $\{ a_{N+1}^f,a_{N+2}^f, a_{N+3}^f, a_{N+4}^f\}$ (in the case of $f_+$, $g_+$, $h_\perp$, and $\tilde{h}_\perp$) and  $\{ a_{N+1}^f,a_{N+2}^f, a_{N+3}^f\}$ (in the case of the other form factors) in terms of the lower-order coefficients $\{a_0^f,a_1^f,...,a_N^f\}$; correspondingly in Eqs.~(\ref{eq:FFfit}) and (\ref{eq:zphys}), we set $M=4$ for  $f_+$, $g_+$, $h_\perp$, and $\tilde{h}_\perp$, and $M=3$ for the other form factors. The explicit expressions for the eliminated higher-order coefficients are given in Appendix \ref{sec:asymptotic}.

Unlike in earlier determinations of $\Lambda_b$-decay form factors \cite{Detmold:2015aaa, Detmold:2016pkz}, here we do not perform separate ``nominal'' and ``higher-order'' fits, and instead directly use a ``higher-order'' fit to obtain the physical-limit form factors. This has the advantage that the resulting covariance matrix of the fit parameters directly gives the total (statistical plus systematic) uncertainties. Moreover, instead of truncating the $z$-expansion at $N=2$, with ad-hoc priors on the sizes of the second-order coefficients $a_2^f$, we now use dispersive bounds on the full set of $z$-expansion coefficients \cite{Boyd:1994tt,Blake:2022vfl,Flynn:2023qmi}, and increase $N$ until the central values and uncertainties of the physical-limit form factors stabilize in the full semileptonic region $0\leq q^2\leq q^2_{\rm max}$. Our final results use $N=5$. More details on this procedure are given farther below.

Among the parameters describing the pion-mass and lattice-spacing dependence, $c_0^f$, $c_1^f$, $b^f$, and $d^f$ were left unconstrained, while the higher-order parameters $\tilde{c}_0^f$, $\tilde{c}_1^f$, $\tilde{b}^f$, $\hat{d}^f$, $\tilde{d}^f$, and $j^f$, which are not needed to describe the data, were constrained with Gaussian priors with central values and widths as in Eqs.~(40)-(48) of Ref.~\cite{Detmold:2016pkz}. By including these higher-order terms with priors limiting them to be not unnaturally large, systematic uncertainties from such higher-order effects are incorporated in the final form-factor results. Note that the $m_\pi^3$ terms are nonanalytic in the quark mass; such terms can arise from loop corrections in chiral perturbation theory. We do not separately include $m_\pi^2\log(m_\pi^2/\mu^2)$ terms, because their dependence on the quark mass can be reasonably well mimicked by the combination of $m_\pi^2$ and $m_\pi^3$ terms in the range considered.

The heavy-quark contributions to the discretization errors are expected to have a nontrivial dependence on the heavy-quark action parameters that is not fully captured by Eq.~(\ref{eq:FFfit}). The size of these discretization errors can be estimated using Symanzik effective theory. As discussed in Appendix \ref{sec:HQdisc}, we expect that residual heavy-quark discretization errors that remain after continuum extrapolation are negligible compared to the other uncertainties.

Besides the effects of the higher-order pion-mass and lattice-spacing terms, our fit also incorporates the following sources of systematic uncertainty:
\begin{enumerate}
    \item When renormalizing the vector and axial-vector currents, the bootstrap samples for the quantities $R_f$  [e.g., Eq.~(\ref{eq:Rhplus})] were generated with the residual matching factors and the $\mathcal{O}(a)$-improvement coefficients drawn from Gaussian random distributions, with central values and widths given in Table~\ref{tab:Pmatchingfactors}. Furthermore, in the data covariance matrix used for the fit, we included an additional $1\%$ renormalization uncertainty for both the vector and axial-vector form factors to more conservatively account for missing higher-order corrections to the residual matching factors $\rho_{\Gamma}$; this uncertainty is taken to be 100\% correlated between either the vector or axial-vector form factors.
    \item For the tensor form factors, the renormalization uncertainty is dominated by the use of the tree-level values, $\rho_{T^{\mu\nu}}=1$, for the residual matching factors in the mostly nonperturbative renormalization procedure. Following Ref.~\cite{Detmold:2016pkz}, we estimate the systematic uncertainty in $\rho_{T^{\mu\nu}}$ to be equal to 2 times the maximum value of $|\rho_{V^{\mu}}-1|$, $|\rho_{A^{\mu}}-1|$, which is equal to $0.05316$. Note that $\rho_{T^{\mu\nu}}$ for the tensor current is scale-dependent, and our estimate of the matching uncertainty (and the values of the form factors themselves) should be interpreted as corresponding to $\mu=4.2$ GeV. To incorporate the tensor-current matching uncertainty in the fit, we introduced a single common nuisance parameter multiplying all tensor form factors on all ensembles, with a Gaussian prior equal to $1\pm 0.05316$.
    \item In Refs.~\cite{Detmold:2015aaa,Meinel:2017ggx}, the finite-volume errors in the $\Lambda_b \to N$ and $\Lambda_c \to N$ form factors were estimated to be 3\% for the smallest value of $m_\pi L$ used there. It is reasonable to expect similar behavior for the $\Xi_b\to\Xi$ form factors. However, the smallest value of $m_\pi L$ is larger here, and we therefore rescale the above estimate with the exponential of the ratio of the smallest $m_\pi L$, leading to an estimate of 1\%. The isospin breaking effects are again estimated to be $\mathcal{O}((m_d-m_u)/\Lambda_{\rm QCD}) \approx 0.5\%$, and $\mathcal{O}(\alpha_{\rm e.m.}) \approx 0.7\%$. The finite-volume and isospin-breaking uncertainties were added to the data covariance matrix used in the fit, assuming $100\%$ correlation within either the vector, axial-vector, and tensor form factors.
    \item The uncertainties in the lattice spacings and pion masses were included by promoting these values to fit parameters with Gaussian priors determined by the central values and uncertainties listed in Table~\ref{tab:lattice_params}.

\end{enumerate}
This calculation uses 2+1-flavor gauge configurations and omits the effects of charm (and heavier) sea quarks. For the hadronic quantities that are determined most precisely from lattice QCD, such as the kaon decay constant and the kaon semileptonic form factor, no significant differences are seen between the FLAG averages of 2+1 and 2+1+1-flavor results, at the level of $< 0.5\%$ uncertainty \cite{FlavourLatticeAveragingGroupFLAG:2024oxs}. We therefore expect the systematic errors  due to the omission of heavier sea quarks to be negligible at our current level of precision.

\begin{table}
\begin{tabular}{ccccc}
\hline\hline
 $f$     & \hspace{1ex} & $J^P$ & \hspace{1ex}  & $m_{\rm pole}^f$ [GeV]  \\
\hline
$f_+$, $f_\perp$, $h_+$, $h_\perp$,                        && $1^-$   && $5.416$  \\
$f_0$                                                      && $0^+$   && $5.711$  \\
$g_+$, $g_\perp$, $\tilde{h}_+$, $\tilde{h}_\perp$,        && $1^+$   && $5.750$  \\
$g_0$                                                      && $0^-$   && $5.367$  \\
\hline\hline
\end{tabular}
\caption{\label{tab:polemasses}$B_s$ meson poles in each of the different form factors.}
\end{table}

For the parametrization (\ref{eq:zphys}), the dispersive bound takes the form\footnote{We do not include the one-particle contributions due to the $B_s$ bound states here, because some of the decay constants are not well known. By including these contributions, the bounds could be tightened slightly. An exploratory analysis including these contributions, using $f_{B_s}$ from Ref.~\cite{FlavourLatticeAveragingGroupFLAG:2024oxs} together with preliminary results for the positive-parity decay constants $f_{B_{s0}^*}$, $f_{B_{s1}}$ from Ref.~\cite{Guyton2026}, and approximating $f^T_{B_s^*}\sim f_{B_s}^*\sim f_{B_s}$, $f^T_{B_{s1}}\sim f_{B_{s1}}$, shows that the uncertainties in our results for the $\Xi_b\to \Xi$ form factors at $q^2=0$ are reduced by $\approx 0.05\sigma$, and the changes in the central values are negligibly small. } \cite{Blake:2022vfl,Flynn:2023qmi}
\begin{eqnarray}
D_{f_0}\leq 1,\: D_{f_{+,\perp}}\leq 1,\:D_{g_0}\leq 1,\: D_{g_{+,\perp}}\leq 1,\: D_{h_{+,\perp}}\leq 1,\:D_{\tilde{h}_{+,\perp}}\leq 1, \label{eq:DB}
\end{eqnarray}
where
\begin{eqnarray}
D_{f_0} &=& \sum_{n=0}^{N+3}\sum_{n^\prime=0}^{N+3} a_n^{f_0} \langle z^n|z^{n^\prime}\rangle_{\alpha_V} a_{n'}^{f_0}, \\
D_{f_{+,\perp}} &=& \sum_{n=0}^{N+4}\sum_{n^\prime=0}^{N+4} a_n^{f_+} \langle z^n|z^{n^\prime}\rangle_{\alpha_V} a_{n'}^{f_+} +  \sum_{n=0}^{N+3}\sum_{n^\prime=0}^{N+3} a_n^{f_\perp} \langle z^n|z^{n^\prime}\rangle_{\alpha_V} a_{n'}^{f_\perp}, 
\end{eqnarray}

\begin{eqnarray}
D_{g_0} &=& \sum_{n=0}^{N+3}\sum_{n^\prime=0}^{N+3} a_n^{g_0} \langle z^n|z^{n^\prime}\rangle_{\alpha_A} a_{n'}^{g_0}, \\
D_{g_{+,\perp}} &=& \sum_{n=0}^{N+4}\sum_{n^\prime=0}^{N+4} a_n^{g_+} \langle z^n|z^{n^\prime}\rangle_{\alpha_A} a_{n'}^{g_+} +  \sum_{n=0}^{N+3}\sum_{n^\prime=0}^{N+3} a_n^{g_\perp} \langle z^n|z^{n^\prime}\rangle_{\alpha_V} a_{n'}^{g_\perp}, 
\end{eqnarray}

\begin{eqnarray}
D_{h_{+,\perp}} &=& \sum_{n=0}^{N+3}\sum_{n^\prime=0}^{N+3} a_n^{h_+} \langle z^n|z^{n^\prime}\rangle_{\alpha_V} a_{n'}^{h_+} +  \sum_{n=0}^{N+4}\sum_{n^\prime=0}^{N+4} a_n^{h_\perp} \langle z^n|z^{n^\prime}\rangle_{\alpha_V} a_{n'}^{h_\perp}, \\
D_{\tilde{h}_{+,\perp}} &=& \sum_{n=0}^{N+3}\sum_{n^\prime=0}^{N+3} a_n^{\tilde{h}_+} \langle z^n|z^{n^\prime}\rangle_{\alpha_A} a_{n'}^{\tilde{h}_+} +  \sum_{n=0}^{N+4}\sum_{n^\prime=0}^{N+4} a_n^{\tilde{h}_\perp} \langle z^n|z^{n^\prime}\rangle_{\alpha_A} a_{n'}^{\tilde{h}_\perp}, 
\end{eqnarray}
with
\begin{eqnarray}
\langle z^n|z^{n^\prime}\rangle_{\alpha} &=& \left\{ \begin{array}{ll} \displaystyle\frac{\sin(\alpha(n-n^\prime))}{\pi (n-n^\prime)}, & n\neq n^\prime, \\ \displaystyle\frac{\pi}{\alpha}, & n=n^\prime, \end{array} \right.
\end{eqnarray}
and
\begin{eqnarray}
\alpha_V &=& \text{arg}\:{z((m_{\Xi_b} + m_{\Xi})^2,t_0,(m_B+m_K)^2)} \:\approx\: 1.43829, \\
\alpha_A &=& \text{arg}\:{z((m_{\Xi_b} + m_{\Xi})^2,t_0,(m_{B^*}+m_K)^2)} \:\approx\: 1.47256.
\end{eqnarray}
To implement the dispersive bounds, we use the algorithm proposed in Ref.~\cite{Flynn:2023qmi}, which we apply directly to the fit of the lattice data using Eq.~(\ref{eq:FFfit}):

\begin{enumerate}
\item We perform a fit using Eq.~(\ref{eq:FFfit}) in which the following term is added to the $\chi^2$ function:
\begin{equation}
\left(D_{f_0}+D_{f_{+,\perp}}+D_{g_0}+D_{g_{+,\perp}}+D_{h_{+,\perp}}+D_{\tilde{h}_{+,\perp}}\right)/\sigma_D^2,
\end{equation}
where $\sigma_D$ is an arbitrary width of order 1 that can be tuned to optimize the efficiency of the algorithm (we use $\sigma_D=0.5$).
\item We generate a large number ($\sim$ $10^6$ to $10^{10}$, depending on $N$) of multivariate Gaussian random samples for the $10 (N+1)-2$ $z$-expansion coefficients that are actual fit parameters, with mean given by the best fit point and covariance matrix computed using the Hessian of $\chi^2$ at the best-fit point.
\item For each sample, we check whether the dispersive bound (\ref{eq:DB}) is satisfied [this requires computing the derived coefficients $a_0^{g_\perp}$, $a_0^{\tilde{h}_\perp}$, and $\{a_{N+1}^f,...,a_{N+M}^f$\} using Eqs.~(\ref{eq:a0gperp}), (\ref{eq:a0htildeperp}), (\ref{eq:M4j}-\ref{eq:M4jp3}), (\ref{eq:M3j}-\ref{eq:M3jp2})]. If not, the sample is discarded. If yes, we draw a single uniformly distributed random number $p\in[0,1]$. We then accept the sample if
\begin{eqnarray}
p&\leq& \frac{\exp\left[-3/\sigma_D^2\right]}{\exp\left[\left(D_{f_0}+D_{f_{+,\perp}}+D_{g_0}+D_{g_{+,\perp}}+D_{h_{+,\perp}}+D_{\tilde{h}_{+,\perp}}\right)/(2\sigma_D^2)\right]}.
\end{eqnarray}

\end{enumerate}

One can then compute the observable of interest (such as the value of a form factor, or a decay rate depending on the form factors) for each of the accepted samples to obtain the probability distribution for the observable. We have done this for the values of the form factors at $q^2=0$ and $q^2=q^2_{\rm max}$ and found that the distributions are reasonably close to Gaussian, and it is sufficient to instead evaluate the form factors using the mean vector of the accepted parameter samples, and evaluate their uncertainty with derivative-based error propagation using the covariance matrix of the $z$-expansion parameters computed from the accepted samples. We have performed the analysis for different $z$-expansion orders up to $N=6$, and found that the central values and uncertainties of the form factors at $q^2=0$ (which, in the semileptonic region, is the point with the highest sensitivity to $N$, as it corresponds to the farthest extrapolation of the lattice data) stabilize at $N=5$ and are practically unchanged when going to $N=6$. Therefore, we choose the $N=5$ results as our final results.

We tested the convergence of the means, standard deviations, and correlations of the accepted samples with the sample size by successively increasing the number of initial samples by factors of 4 until the changes in the means and standard deviations of the accepted samples are below $0.015$ times the standard deviations for all parameters, and the absolute changes in all elements of the correlation matrix are below $0.025$. The predictions for the form factors and for the physical observables considered in this work were seen to stabilize well before these criteria were reached.

In the Supplemental Material \cite{Supplemental}, we provide machine-readable files containing, for the 58 $z$-expansion fit parameters,
\begin{enumerate}
\item The 68446 accepted samples (out of 320,000,000 initial samples) from the above procedure,
\item The mean vector and covariance matrix of the accepted samples,
\item The best-fit values and covariance matrix from the initial fit in step 1 above. These are included for reproducibility and possibly different statistical processing, and should not be used directly in phenomenological applications.
\end{enumerate}
The mean vector and standard deviations of the accepted samples are also listed in Table~\ref{tab:zfit_params}. Plots of the chiral-continuum extrapolations of the form factors are shown in Figs.~\ref{fig:vector_extrap}, \ref{fig:axial_vector_extrap}, \ref{fig:tensor_extrap}, and \ref{fig:tensor_tilde_extrap}. The physical-limit curves were evaluated using the mean vector and covariance matrix of the accepted samples.

\begin{table}
\begin{tabular}{ccccccc}
\hline \hline
$f$  & $a_0^f$ & $a_1^f$ & $a_2^f$ & $a_3^f$ & $a_4^f$ & $a_5^f$  \\
\hline
$f_0$  & $ 0.0749\pm 0.0027$ & $-0.573\pm 0.063$ & $ 2.89\pm 0.76$ & $-8.9\pm 4.3$ & $ 16\pm 11$ & $-19\pm 16$ \\ 
$f_+$  & $ 0.0307\pm 0.0010$ & $-0.265\pm 0.022$ & $ 1.33\pm 0.31$ & $-4.1\pm 2.3$ & $  8\pm  8$ & $-10\pm 14$ \\ 
$f_\perp$  & $ 0.0771\pm 0.0025$ & $-0.570\pm 0.050$ & $ 2.33\pm 0.58$ & $-5.6\pm 3.6$ & $  8\pm 10$ & $ -9\pm 14$ \\ 
$g_0$  & $ 0.0899\pm 0.0035$ & $-0.699\pm 0.078$ & $ 3.06\pm 0.81$ & $-7.1\pm 4.2$ & $ 10\pm 11$ & $ -9\pm 15$ \\ 
$g_+$  & $ 0.0185\pm 0.0006$ & $-0.148\pm 0.014$ & $ 0.73\pm 0.21$ & $-2.2\pm 1.6$ & $  4\pm  5$ & $ -6\pm  9$ \\ 
$g_\perp$  &                       & $-0.412\pm 0.039$ & $ 1.95\pm 0.51$ & $-4.4\pm 3.2$ & $  6\pm  8$ & $ -5\pm 12$ \\ 
$h_+$  & $ 0.0631\pm 0.0040$ & $-0.387\pm 0.047$ & $ 1.02\pm 0.52$ & $-1.7\pm 3.4$ & $  2\pm  9$ & $ -2\pm 14$ \\ 
$h_\perp$  & $ 0.0468\pm 0.0030$ & $-0.368\pm 0.042$ & $ 1.52\pm 0.47$ & $-3.5\pm 3.3$ & $  5\pm 11$ & $ -5\pm 19$ \\ 
$\widetilde{h}_+$  & $ 0.0469\pm 0.0030$ & $-0.332\pm 0.037$ & $ 1.58\pm 0.45$ & $-4.1\pm 3.0$ & $  6\pm  8$ & $ -7\pm 12$ \\ 
$\widetilde{h}_\perp$  &                       & $-0.251\pm 0.026$ & $ 1.33\pm 0.35$ & $-4.8\pm 2.6$ & $ 11\pm  8$ & $-16\pm 15$ \\ \hline\hline
\end{tabular}
\caption{\label{tab:zfit_params} Mean values and standard deviations of the accepted samples for the $z$-expansion parameters. Machine-readable files containing these values and their covariance matrix with more digits are provided in the supplemental material \cite{Supplemental}. The parameters $a_0^{g_\perp}$ and $a_0^{\tilde{h}_\perp}$, as well as $\{a_6^f,a_7^f,a_8^f,a_9^f$\} for $f\in\{f_+, g_+, h_\perp, \tilde{h}_\perp\}$ and $\{a_6^f,a_7^f,a_8^f$\} for $f\in\{f_0,f_\perp g_0,g_\perp,h_+,\tilde{h}_+\}$, must be determined from these parameters using Eqs.~(\protect\ref{eq:a0gperp}), (\protect\ref{eq:a0htildeperp}), (\protect\ref{eq:M4j}-\protect\ref{eq:M4jp3}), (\protect\ref{eq:M3j}-\protect\ref{eq:M3jp2}).}
\end{table}

\begin{figure}
    \centering
    \hspace{1cm}
    \includegraphics[width=0.8\linewidth]{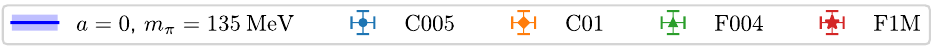}
    
    \includegraphics[width=\linewidth]{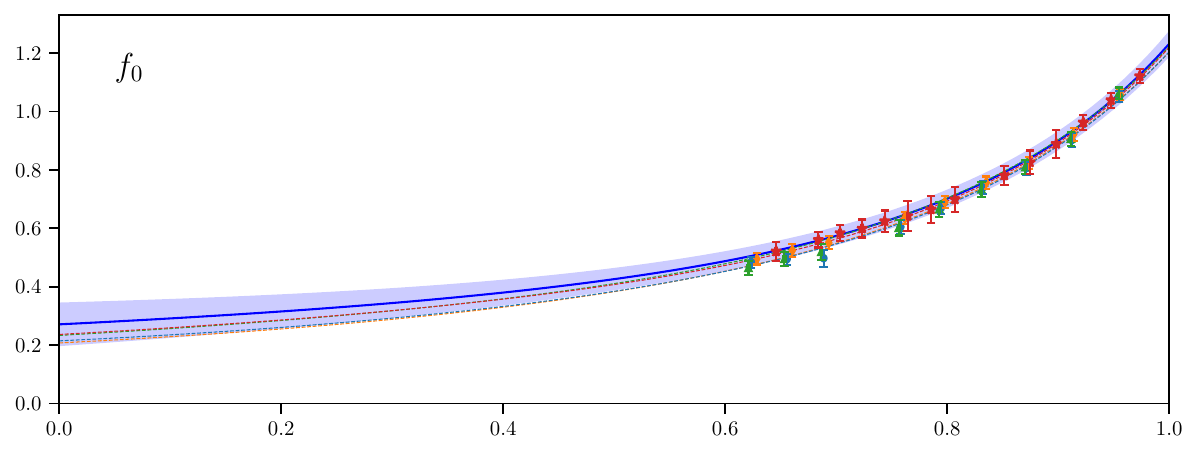}

    \includegraphics[width=\linewidth]{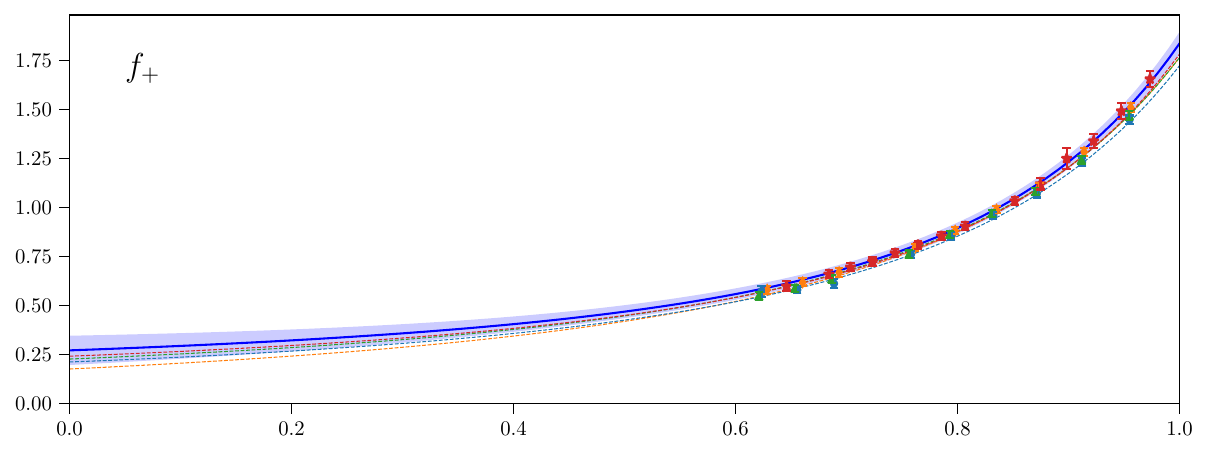}

    \includegraphics[width=\linewidth]{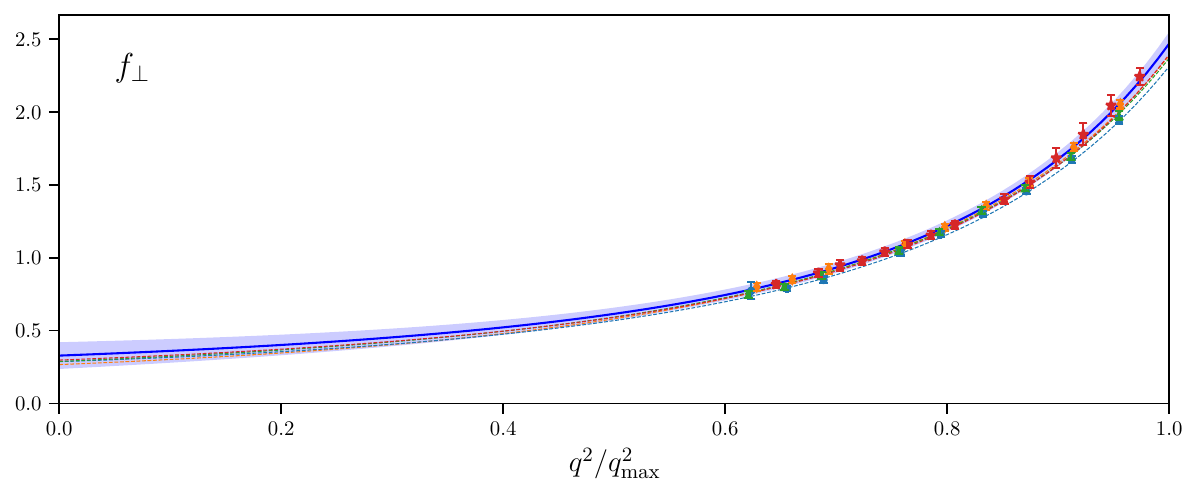}

    \caption{Chiral and continuum extrapolations of the $\Xi_b \to \Xi$ vector form factors. The solid blue lines show the form factor curves in the physical limit, while the dashed lines show the modified $z$-expansion fits evaluated with the individual lattice spacings and pion masses for each ensemble. The bands include the combined statistical and systematic uncertainties.}
    \label{fig:vector_extrap}
\end{figure}

\begin{figure}
    \centering
    \hspace{1cm}
    \includegraphics[width=0.8\linewidth]{figs/ff_legend.pdf}
    
    \includegraphics[width=\linewidth]{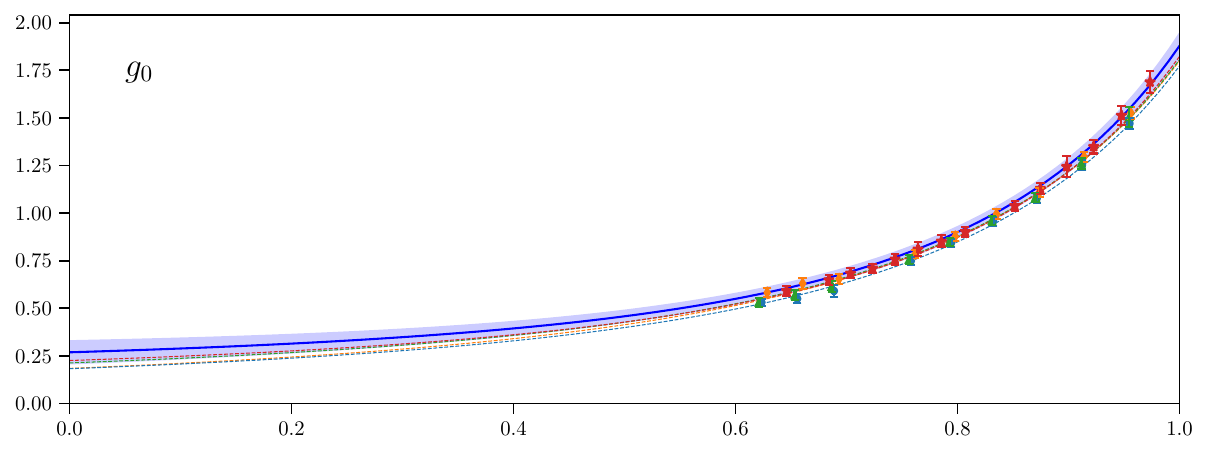}

    \includegraphics[width=\linewidth]{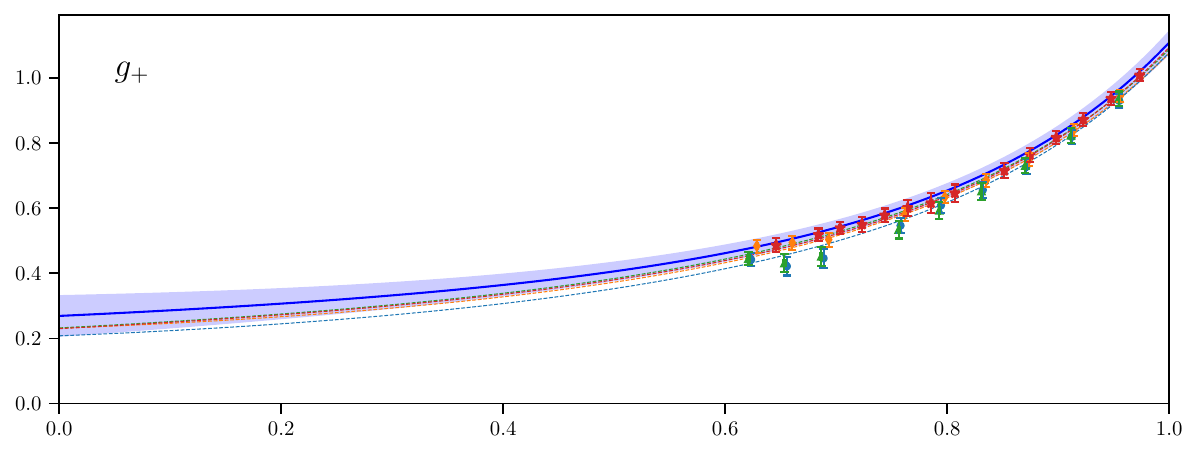}

    \includegraphics[width=\linewidth]{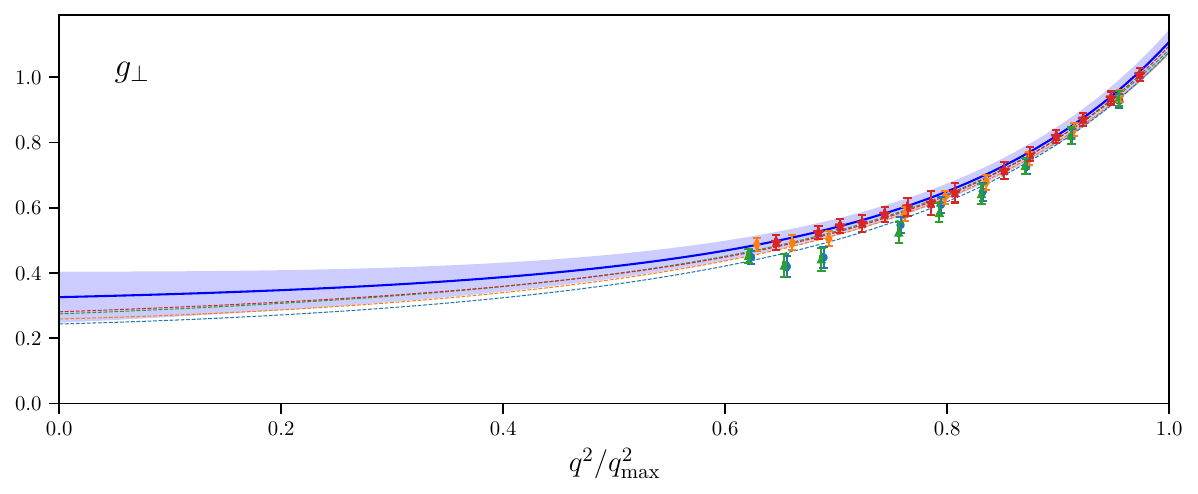}

    \caption{Like Fig.~\ref{fig:vector_extrap}, but for the axial-vector form factors.}
    \label{fig:axial_vector_extrap}
\end{figure}

\begin{figure}
    \centering
    \hspace{1cm}
    \includegraphics[width=0.8\linewidth]{figs/ff_legend.pdf}

    \includegraphics[width=\linewidth]{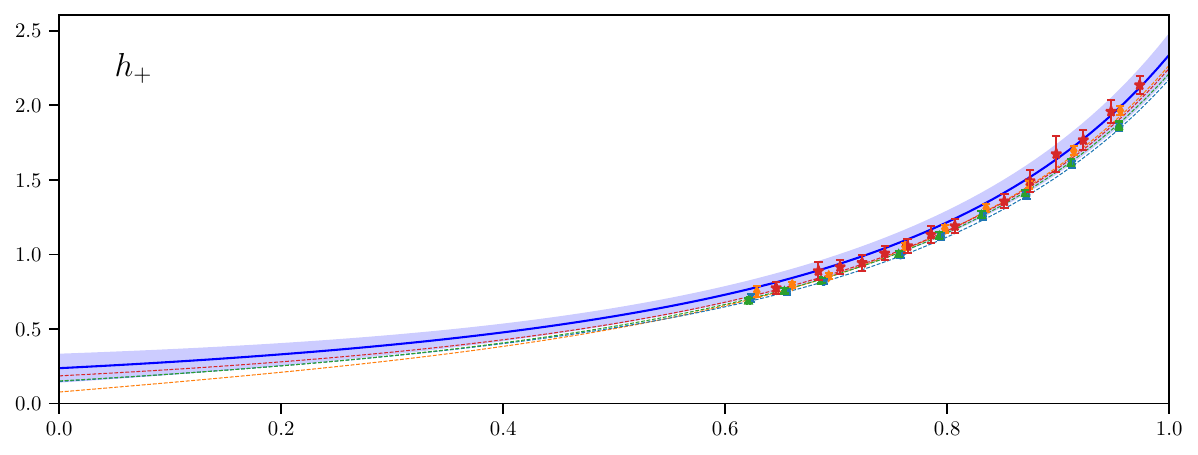}

    \includegraphics[width=\linewidth]{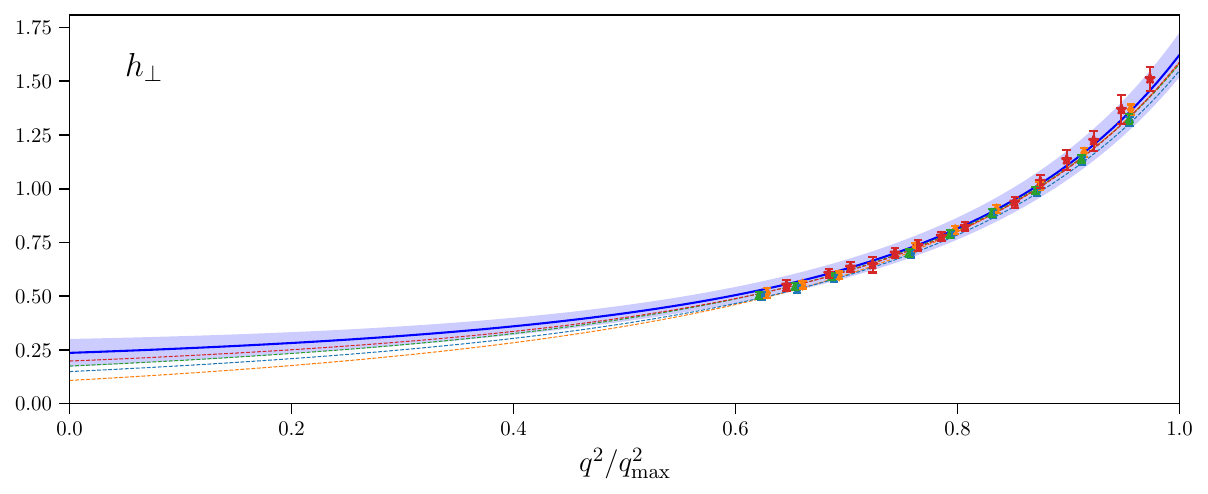}

    \caption{Like Fig.~\ref{fig:vector_extrap}, but for the tensor form factors $h_+$ and $h_\perp$.}
    \label{fig:tensor_extrap}
\end{figure}

\begin{figure}
    \centering
    \hspace{1cm}
    \includegraphics[width=0.8\linewidth]{figs/ff_legend.pdf}

    \includegraphics[width=\linewidth]{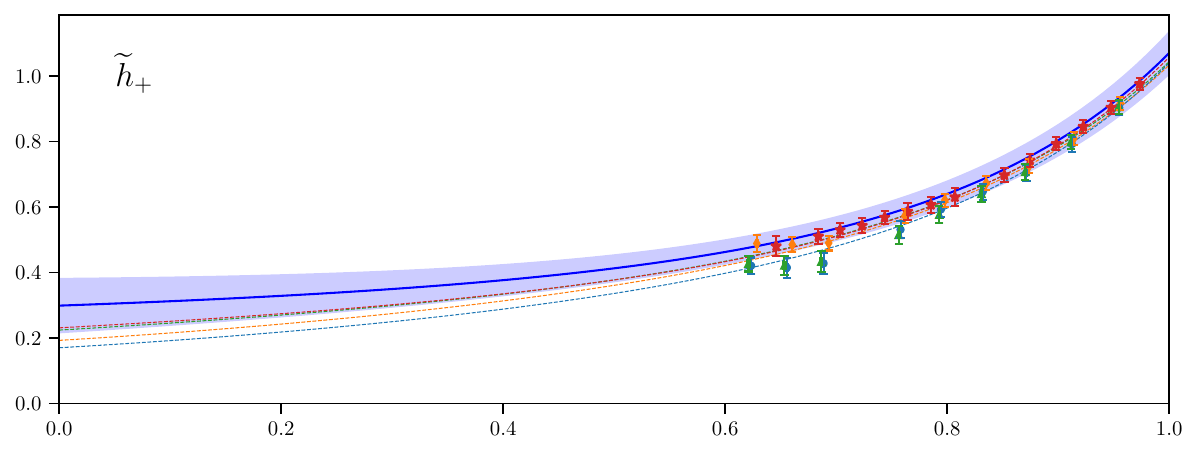}

    \includegraphics[width=\linewidth]{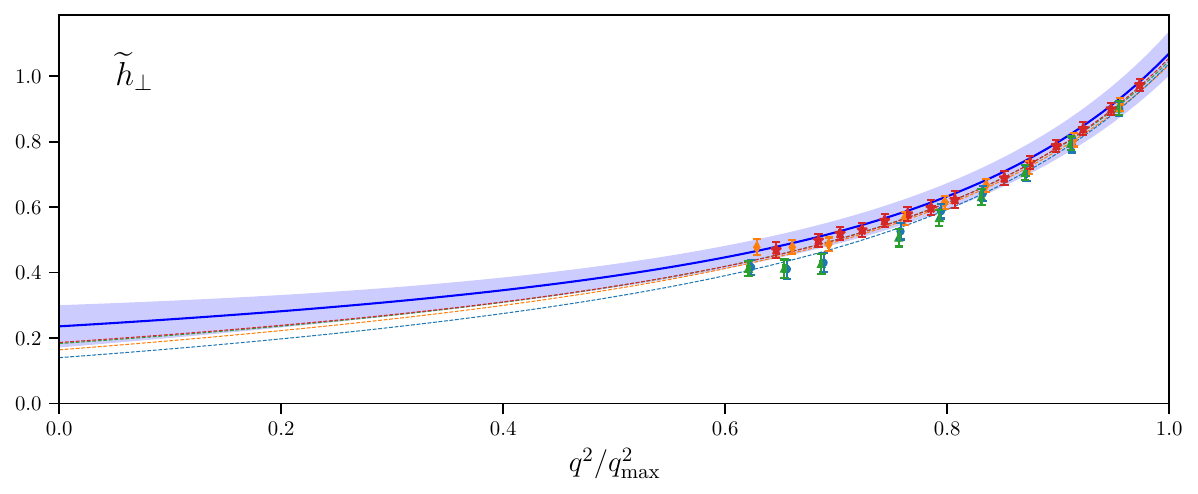}

    \caption{Like Fig.~\ref{fig:vector_extrap}, but for the tensor form factors  $\tilde{h}_+$ and $\tilde{h}_\perp$.}
    \label{fig:tensor_tilde_extrap}
\end{figure}

\FloatBarrier
\section{Standard-Model Predictions for \texorpdfstring{$\bm{\Xi^-_b \to \Xi^- \mu^+\mu^-}$}{Xib(-) to Xi(-) mu(+) mu(-)} and \texorpdfstring{$\bm{\Xi_b^- \to \Xi^- \gamma}$}{Xib(-) to Xi(-) gamma} decays}
\FloatBarrier

\begin{table}
\begin{tabular}{lcccccc}
\hline\hline
            & & $\mu=2.1\:{\rm GeV}$ & & $\mu=4.2\:{\rm GeV}$ & & $\mu=8.4\:{\rm GeV}$ \\
\hline
$C_1$       && $-0.4996$     && $-0.2891$      && $-0.1494$      \\
$C_2$       && $\wm1.0248$   && $\wm1.0102$    && $\wm1.0036$    \\
$C_3$       && $-0.0145$     && $-0.0061$      && $-0.0027$      \\
$C_4$       && $-0.1517$     && $-0.0866$      && $-0.0546$      \\
$C_5$       && $\wm0.0010$   && $\wm0.0004$    && $\wm0.0002$    \\
$C_6$       && $\wm0.0033$   && $\wm0.0011$    && $\wm0.0004$    \\
$C_7$       && $-0.3786$     && $-0.3362$      && $-0.3035$      \\
$C_8$       && $-0.2133$     && $-0.1823$      && $-0.1630$      \\
$C_9$       && $\wm4.5589$   && $\wm4.2665$    && $\wm3.8628$    \\
$C_{10}$    && $-4.1333$     && $-4.1133$      && $-4.1333$      \\
$m_b^{\overline{\rm MS}}\:[{\rm GeV}]$
            && $\wm4.9021$   && $\wm4.1737$    &&  $\wm3.7239$   \\
$\alpha_s$  && $\wm0.2972$   && $\wm0.2248$    &&  $\wm0.1860$   \\
$\alpha_e$  && $\wm1/134.44$ && $\wm1/133.28$  &&  $\wm1/132.51$ \\
\hline\hline
\end{tabular}
\caption{\label{tab:Wilson}Wilson coefficients, $b$-quark mass, and strong and electromagnetic couplings in the $\overline{\rm MS}$ scheme at the nominal scale $\mu=4.2\:{\rm GeV}$ and at the low and high scales
used to estimate the perturbative uncertainties.}
\end{table}

\begin{figure}
\includegraphics[height=0.25\textheight]{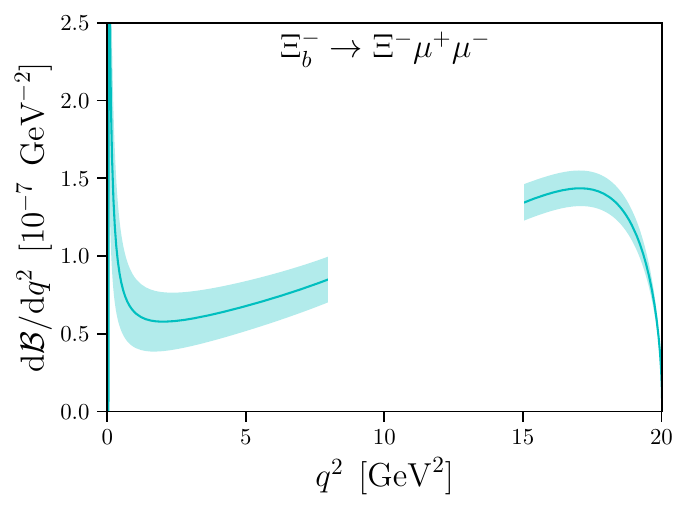}

\includegraphics[height=0.25\textheight]{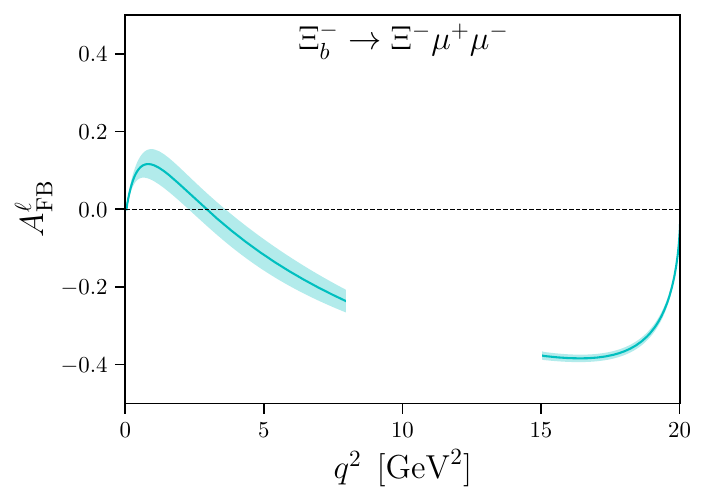} 

\includegraphics[height=0.25\textheight]{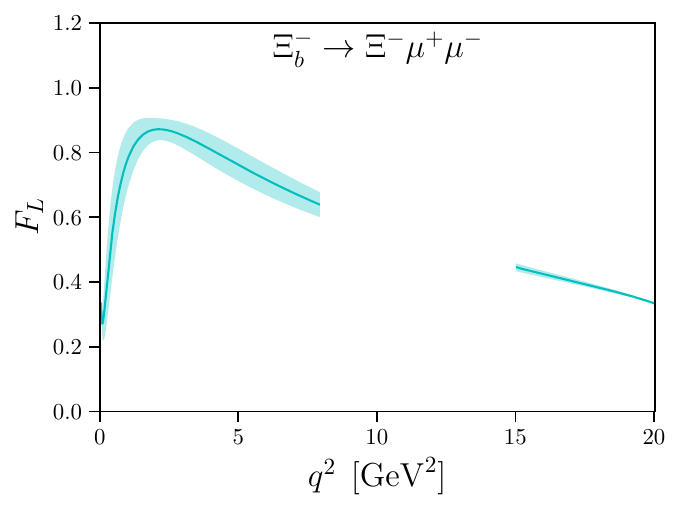}
\caption{\label{fig:XibXimmumu}Standard-Model predictions for the $\Xi^-_b \to \Xi^- \mu^+\mu^-$ differential branching fraction and the angular observables $F_L$ and $A_{FB}^\ell$.}
\end{figure}

To predict the $\Xi^-_b \to \Xi^- \mu^+\mu^-$ differential branching fraction and angular observables in the Standard Model, we closely follow Ref.~\cite{Detmold:2016pkz} (and references therein), which considered $\Lambda_b \to \Lambda \mu^+\mu^-$. Here, we do not include the secondary decay of the $\Xi^-$ and calculate only the two-fold decay distribution
\begin{equation}
\frac{\mathrm{d}^2\Gamma}{\mathrm{d}q^2 \mathrm{d}\cos\theta_\ell}=\frac{3}{2}\left( K_{1ss} \sin^2\theta_\ell +\, K_{1cc} \cos^2\theta_\ell + K_{1c} \cos\theta_\ell \right),
\end{equation}
where $\theta_\ell$ is the polar angle of the negatively charged lepton \cite{Boer:2014kda}. The integral over $\cos\theta_\ell$ gives the $q^2$-differential decay rate
\begin{equation}
\frac{\mathrm{d}\Gamma}{\mathrm{d}q^2}=2 K_{1ss} + K_{1cc}.
\end{equation}
The two specific angular observables we consider are
\begin{eqnarray}
F_L &=& 2 \hat{K}_{1ss} - \hat{K}_{1cc}, \\
A_{FB}^\ell &=& \frac{3}{2} \hat{K}_{1c},
\end{eqnarray}
where $\hat{K}_i=K_i/(\mathrm{d}\Gamma/\mathrm{d}q^2)$. We limit our predictions to the kinematic regions $q^2<8\:{\rm GeV}^2$ and $q^2>15\:{\rm GeV}^2$ to avoid the effects of narrow charmonium resonances, and use the approximation in which all nonlocal hadronic matrix elements are reduced to local matrix elements through the effective Wilson coefficients $C_7^{\rm eff}(q^2)$ and $C_9^{\rm eff}(q^2)$, given in Eqs.~(65) and (66) of Ref.~\cite{Detmold:2016pkz} (following  Ref.~\cite{Du:2015tda}), which are based on Refs.~\cite{Beneke:2001at,Asatryan:2001zw,Grinstein:2004vb,Greub:2008cy,Beylich:2011aq}. We take the Standard-Model expressions for the functions $K_i$ in this approximation, and including lepton-mass effects, from Eqs.~(A1) and (A2) of Ref.~\cite{Gutsche:2013pp} (with the appropriate replacements of the baryon masses, and using the form-factor relations given in Ref.~\cite{Detmold:2016pkz}). To evaluate the $\Xi_b\to\Xi$ form factors, we use the mean values and covariance matrix of the accepted samples (cf.~Sec.~\ref{sec:extrap}) of the $z$-expansion coefficients.  The values we use for the Wilson coefficients $C_1$-$C_{10}$, the $b$-quark mass, and the strong and electromagnetic couplings are given in Table \ref{tab:Wilson}; the pole masses that enter in $C_7^{\rm eff}(q^2)$ and $C_9^{\rm eff}(q^2)$ are
\begin{eqnarray}
 m_c^{\rm pole} &=& 1.6030\:\:{\rm GeV}, \\
 m_b^{\rm pole} &=& 4.7231\:\:{\rm GeV}.
\end{eqnarray}
We evaluated all of these as in Ref.~\cite{Detmold:2016pkz}, and, except for $\alpha_e$, updated them using EOS \cite{EOSAuthors:2021xpv} with the latest Standard-Model input parameters from Ref.~\cite{ParticleDataGroup:2024cfk}. We further take
\begin{equation}
|V_{ts}^* V_{tb}| = 0.04113 \pm 0.00036
\label{eq:VtsVtb}
\end{equation}
from UTFit \cite{UTfit} and, to evaluate the differential branching fraction $\mathrm{d}\mathcal{B}/\mathrm{d}q^2=\tau_{\Xi_b^-} \mathrm{d}\Gamma/\mathrm{d}q^2$,
\begin{equation}
 \tau_{\Xi_b^-} = (1.570 \pm 0.023)\:\:{\rm ps} \label{eq:tauXib}
\end{equation}
from Ref.~\cite{ParticleDataGroup:2024cfk}. The resulting Standard-Model predictions are shown as a function of $q^2$ in Fig.~\ref{fig:XibXimmumu}; in addition, we provide predictions for selected $q^2$ bins in Table \ref{tab:binnedobs}. For the angular observables, the numerator and denominator of $\hat{K}_i=K_i/(\mathrm{d}\Gamma/\mathrm{d}q^2)$ are binned first before taking the ratio.

\begin{table}
\begin{tabular}{llll}
\hline \hline
              & $\wm\langle \mathrm{d}\mathcal{B}/\mathrm{d} q^2\rangle$ & $\wm\langle F_L \rangle $ & $\wm\langle A_{\rm FB}^\ell \rangle $   \\
\hline
$[0.1, 2]$    & $\wm0.76(27)$  & $\wm0.65(12)$  & $\wm0.088(25)$  \\ 
$[1.1, 6]$    & $\wm0.63(17)$  & $\wm0.810(42)$  & $-0.038(40)$   \\ 
$[2, 4]$      & $\wm0.60(17)$  & $\wm0.848(39)$  & $-0.004(42)$  \\ 
$[4, 6]$      & $\wm0.68(16)$  & $\wm0.761(49)$  & $-0.117(40)$   \\ 
$[6, 8]$      & $\wm0.79(15)$  & $\wm0.674(43)$  & $-0.203(33)$   \\ 
$[15, 16]$    & $\wm1.375(98)$  & $\wm0.435(11)$  & $-0.380(10)$  \\ 
$[16, 18]$    & $\wm1.424(93)$  & $\wm0.404(83)$  & $-0.3795(94)$  \\ 
$[18, 20]$    & $\wm1.080(72)$  & $\wm0.3635(42)$  & $-0.3089(98)$  \\ 
$[15, 20]$    & $\wm1.277(83)$  & $\wm0.3967(73)$  & $-0.3558(93)$  \\ 
\hline \hline
\end{tabular}
\caption{\label{tab:binnedobs}Standard-Model predictions for the binned $\Xi_b^- \to \Xi^-\, \mu^+ \mu^-$ differential branching fraction (in units of $10^{-7}\:\:{\rm GeV}^{-2}$) and angular observables (with unpolarized $\Xi_b$). The first column specifies the bin ranges $[q^2_{\rm min},\: q^2_{\rm max}]$ in units of ${\rm GeV}^2$.}
\end{table}

For the radiative decay $\Xi_b^- \to \Xi^- \gamma$, we again use the approximation in which nonlocal hadronic matrix elements are rewritten as local matrix elements via $C_7^{\rm eff}(q^2=0)$, such that (see, e.g., Ref.~\cite{Gutsche:2013pp})
\begin{equation}
\mathcal{B}(\Xi_b^- \to \Xi^- \gamma)=\tau_{\Xi_b^-}\alpha_e\left(\frac{G_F\, |V_{ts}^* V_{tb}|\,m_b^{\overline{\rm MS}}\,|C_7^{\rm eff}(0)|}{4\pi^2\sqrt{2}}\right)^2\frac{(m_{\Xi_b^-}^2-m_{\Xi^-}^2)^3}{m_{\Xi_b^-}^3}h_\perp(0)^2,
\end{equation}
where we used the endpoint relation (\ref{eq:endptconst3}). For $C_7^{\rm eff}(0)$, we again use Eq.~(65) of Ref.~\cite{Detmold:2016pkz}, which, at $\mu=4.2$ GeV is found to be
\begin{equation}
C_7^{\rm eff}(0) \approx -0.36795-0.01355\:i.
\end{equation}
We obtain the Standard-Model prediction
\begin{equation}
\mathcal{B}(\Xi_b^- \to \Xi^-\gamma) = (2.9 \pm 1.6)\times 10^{-5}.
\end{equation}
For all observables computed in this section, the uncertainties given include the form-factor and parametric uncertainties, the perturbative uncertainties, estimated as the change in the observable when varying the renormalization scale from $\mu=2.1\:{\rm GeV}$ to $\mu=8.4\:{\rm GeV}$, and, following Ref.~\cite{Detmold:2016pkz}, an additional 5\% uncertainty assigned to $C_7^{\rm eff}(q^2)$ and $C_9^{\rm eff}(q^2)$ in an attempt to account for some of the approximations used for the nonlocal matrix elements. Our analysis neglects nonfactorizable effects, whose theory for baryonic decays is only partially developed (see Ref.~\cite{Feldmann:2023plv} for recent progress).

\FloatBarrier
\section{Conclusions}
\FloatBarrier

This work is the first lattice QCD calculation of $\Xi_b \to \Xi$ form factors, and also introduces an important methodological advance in lattice QCD calculations of $b$-baryon decay form factors: the use of dispersive bounds in the chiral-continuum-kinematic extrapolations, which has allowed us to increase the order of the $z$ expansion until the values and uncertainties of the extrapolated form factors stabilize, thereby achieving controlled uncertainties in the full semileptonic kinematic range. In the high-$q^2$ region, our final results in the continuum limit and at the physical pion mass have uncertainties of approximately 3-4\% for the vector and axial-vector form factors, and approximately 6-7\% for the tensor form factors. At $q^2=0$, the uncertainties in all form factors are on the order of 25\%.

If $SU(3)$ flavor symmetry was exact, the $\Xi_b \to \Xi$ form factors would be equal to the $\Lambda_b \to p$ form factors, and equal to $\sqrt{3/2}$ times the $\Lambda_b \to \Lambda$ form factors. A comparison with the QCD lattice results for $\Lambda_b \to p$ \cite{Detmold:2015aaa} and $\Lambda_b \to \Lambda$ \cite{Detmold:2016pkz} shows that these relations are indeed satisfied up to the expected natural size of $SU(3)$ symmetry breaking.

Comparisons of our results for the $\Xi_b \to \Xi$ form factors with continuum calculations are shown in Tables \ref{tab:FFcompare} and \ref{tab:xi_gamma_branching_ratios} at the kinematic points $q^2=q^2_{\rm max}$ and $q^2=0$. Compared to Rui \textit{et al.}, 2025 \cite{Rui:2025ajk}, the central values of our results tend to be lower, but there is a reasonable overall agreement given the estimated uncertainties. A larger number of continuum calculations are available for the tensor form factor at $q^2=0$ that is relevant for the $\Xi_b \to\Xi\gamma$ decay rates, as shown in Table \ref{tab:xi_gamma_branching_ratios}. The quark models of Refs. \cite{Davydov:2022glx} and \cite{Geng:2022xpn} gave lower values for this form factor compared to our lattice calculation, while Refs.~\cite{Rui:2025ajk} and especially Ref.~\cite{Liu:2011ema} gave higher values.

\begin{table}[h]
\begin{tabular}{lclcc}
\hline\hline
                  & This work & \hspace{2ex} & Rui \textit{et al.}, 2025 \cite{Rui:2025ajk}  & Azizi \textit{et al.}, 2011 \cite{Azizi:2011mw}  \\
\hline 
$f_0(q^2_{\rm max})$                                                & $1.232 \pm  0.045$  & & $1.525^{+0.005+0.010+0.000}_{-0.087-0.115-0.063}$ & $-0.060$  \\
$f_+(q^2_{\rm max})$                                                & $1.837 \pm  0.059$  & & $2.391^{+0.058+0.137+0.045}_{-0.054-0.039-0.137}$  & 1.628   \\
$f_\perp(q^2_{\rm max})$                                            & $2.468 \pm 0.080$   & & $3.019^{+0.102+0.076+0.104}_{-0.034-0.156-0.184}$  & 3.667   \\
$g_0(q^2_{\rm max})$                                                &  $1.879 \pm 0.073$  & &  $2.060^{+0.022+0.140+0.213}_{-0.066-0.000-0.370}$ & 2.987   \\
$g_+(q^2_{\rm max})=g_\perp(q^2_{\rm max})$                         & $1.108 \pm 0.037$   & & $1.144^{+0.036+0.134+0.165}_{-0.108-0.000-0.296}$  & 0.825   \\
$h_+(q^2_{\rm max})$                                                & $2.34 \pm 0.15$   & & $3.815^{+0.000+0.000+0.000}_{-0.709-1.391-0.881}$  &    \\
$h_\perp(q^2_{\rm max})$                                            & $1.62 \pm 0.10$   & &  $2.899^{+0.000+0.000+0.000}_{-0.439-0.832-0.584}$ &    \\
$\tilde{h}_+(q^2_{\rm max})=\tilde{h}_\perp(q^2_{\rm max})$         & $1.070 \pm  0.067$  & & $1.181^{+0.035+0.078+0.082}_{-0.002-0.062-0.060}$ &    \\[1ex]
$f_0(0)=f_+(0)$                                                     & $0.271 \pm 0.075$  & &  $0.342^{+0.008+0.013+0.018}_{-0.000-0.013-0.024}$  &  0.142  \\
$f_\perp(0)$                                                        & $0.328 \pm 0.092$   & &  $0.457^{+0.010+0.018+0.023}_{-0.000-0.017-0.031}$  &  0.284  \\
$g_0(0)=g_+(0)$                                                     & $0.269 \pm 0.064$   & & $0.354^{+0.010+0.013+0.015}_{-0.000-0.016-0.020}$  &  0.160  \\
$g_\perp(0)$                                                        & $0.326 \pm  0.079$  & & $0.284^{+0.005+0.012+0.012}_{-0.000-0.010-0.015}$  &  0.120  \\
$h_+(0)$                                                            &  $0.237 \pm 0.097$  & &  $0.429^{+0.033+0.029+0.043}_{-0.000-0.000-0.009}$ &    \\
$\tilde{h}_+(0)$                                                    &  $0.299 \pm 0.085$  & & $0.282^{+0.006+0.010+0.014}_{-0.000-0.009-0.018}$  &    \\
$h_\perp(0)=\tilde{h}_\perp(0)$                                     &  $0.236 \pm 0.065$   & & $0.338^{+0.015+0.009+0.024}_{-0.000-0.000-0.015}$  &  \\
\hline\hline
\end{tabular}
\caption{\label{tab:FFcompare}Comparison of our results for the form factors at $q^2=q^2_{\rm max}$ and $q^2=0$ with those from Refs.~\cite{Rui:2025ajk} and \cite{Azizi:2011mw}.  Reference \cite{Rui:2025ajk} provides separate values for $h_\perp(0)$ and $\tilde{h}_\perp(0)$ which are consistent with each other within the uncertainties; the value shown in the table is $h_\perp(0)$. In Ref.~\cite{Azizi:2011mw}, the matrix elements of the tensor current are written in terms of six form
factors, of which only four are independent. The numerical parameterizations given for the six tensor form factors violate
the exact kinematical relations between these form factors, and therefore we only list the vector and axial-vector form factors from Ref.~\cite{Azizi:2011mw}. }
\end{table}

\begin{table}[h]
    \centering
    \begin{tabular}{l c c c c}
        \hline\hline \\[-3.2ex]
        & Method & $h_\perp(0)=\tilde{h}_\perp(0)$ &  $\mathcal{B}(\Xi_b^- \rightarrow \Xi^- \gamma)$  \\
        \hline \\ [-3.2ex]
        LHCb, 2021   \cite{LHCb:2021hfz} & Experiment & & $< 1.3 \times 10^{-4}$  \\ 
        This work  & Lattice QCD &  $0.236 \pm 0.065$ & $(2.9 \pm 1.6)\times 10^{-5}$ \\
        Rui \textit{et al.}, 2025 \cite{Rui:2025ajk}  & pQCD approach & $0.338^{+0.015+0.009+0.024}_{-0.000-0.000-0.015}$  &  \\
        Aliev \textit{et al.}, 2023 \cite{Aliev:2023mdf} & Light-cone sum rules & $0.31\pm0.04$ & $(4.8 \pm 1.3) \times 10^{-5}$  \\
        Davydov \textit{et al.}, 2022 \cite{Davydov:2022glx} & Relativistic quark model  & $0.144\pm0.007$ & $(0.95 \pm 0.15) \times 10^{-5}$  \\  
        Geng \textit{et al.}, 2022  \cite{Geng:2022xpn} & Light-front quark model & $0.143$ & $(1.1 \pm 0.1) \times 10^{-5}$  \\ 
        Olamaei \textit{et al.}, 2021 \cite{Olamaei:2021eyo} & Light-cone sum rules \cite{Azizi:2011mw} & & $(1.08^{+0.63}_{-0.49}) \times 10^{-5}$ \\
        Wang \textit{et al.}, 2021 \cite{Wang:2020wxn} & Flavor-$SU(3)$ symmetry & & $(1.23 \pm 0.64) \times 10^{-5}$ \\
        Liu \textit{et al.}, 2011 \cite{Liu:2011ema} & Light-cone sum rules & $\approx 0.6$ & $(3.03 \pm 0.10) \times 10^{-4}$ \\
        \hline\hline
    \end{tabular}
    \caption{Comparison of results for the tensor form factor $h_\perp(0)=\tilde{h}_\perp(0)$ and of Standard-Model predictions for the $\Xi_b^- \to \Xi^- \gamma$ branching fraction. Also shown is the experimental upper bound on this branching fraction from LHCb. Note that the Flavor-$SU(3)$-symmetry prediction \cite{Wang:2020wxn} uses the LHCb experimental result for $\mathcal{B}(\Lambda_b \to \Lambda \gamma)$ \cite{LHCb:2019wwi} as an input, and could therefore be directly affected by new physics.}
    \label{tab:xi_gamma_branching_ratios}
\end{table}

Our Standard-Model prediction for $\mathcal{B}(\Xi_b^- \to \Xi^-\gamma)$ is consistent with the upper bound $\mathcal{B}(\Xi_b^- \to \Xi^-\gamma)< 1.3 \times 10^{-4}$ from LHCb \cite{LHCb:2021hfz} and with the previous Standard-Model predictions shown in Table \ref{tab:xi_gamma_branching_ratios}, except for that of Ref.~\cite{Liu:2011ema}, which is about a factor 10 higher. Our Standard-Model prediction of the $\Xi^-_b \to \Xi^- \mu^+\mu^-$ differential branching fraction, shown in Fig.~\ref{fig:XibXimmumu}, has central values approximately 25\% below the prediction of Ref.~\cite{Rui:2025ajk}, but is consistent with  Ref.~\cite{Rui:2025ajk} within the combined uncertainties. We caution that the treatment of charm-loop and nonfactorizable effects in the decay processes is unsatisfactory, and more work is needed to fully control these effects; the primary contribution of this paper is the precise lattice calculation of the local form factors. 

\clearpage

\section*{Acknowledgments}

We are grateful to the RBC and UKQCD Collaborations for making their gauge-field ensembles available. We thank Enrico Lunghi for a discussion on the theory of $b\to s\gamma$ decays. We acknowledge financial support by the U.S. Department of Energy, Office of Science, Office of High Energy Physics under Award Number DE-SC0009913.
This research used resources of the National Energy Research Scientific Computing Center (NERSC), a U.S.~Department of Energy Office of Science User Facility supported by Contract Number DE-AC02-05CH1123. This research also used resources at Purdue University RCAC and at the University of Texas TACC through the Extreme Science and Engineering Discovery Environment (XSEDE) \cite{XSEDE} and the Advanced Cyberinfrastructure Coordination Ecosystem: Services \& Support (ACCESS) program \cite{10.1145/3569951.3597559}, which are supported by U.S. National Science Foundation grants ACI-154856, 2138259, 2138286, 2138307, 2137603, and 2138296.  We acknowledge the use of Chroma \cite{Edwards:2004sx,Chroma}, QLUA \cite{QLUA}, MDWF \cite{MDWF}, and related USQCD software \cite{USQCD}.

\section*{Data Availability}

Machine-readable files containing the accepted samples of the form-factor parameters, their mean and covariance matrix, as well as the central values and covariance matrix from the initial fit prior to the ``reweighting'' procedure, are provided in the Supplemental Material \cite{Supplemental}. The other data are available from the authors upon reasonable request.

\FloatBarrier

\appendix

\section{Additional plots of ratio fits}

This appendix contains sample plots of the ratio fits for the other ensembles (Fig.~\ref{fig:other_ensemble_ratios}).

\begin{figure}
    \centering
    \includegraphics[width=\linewidth]{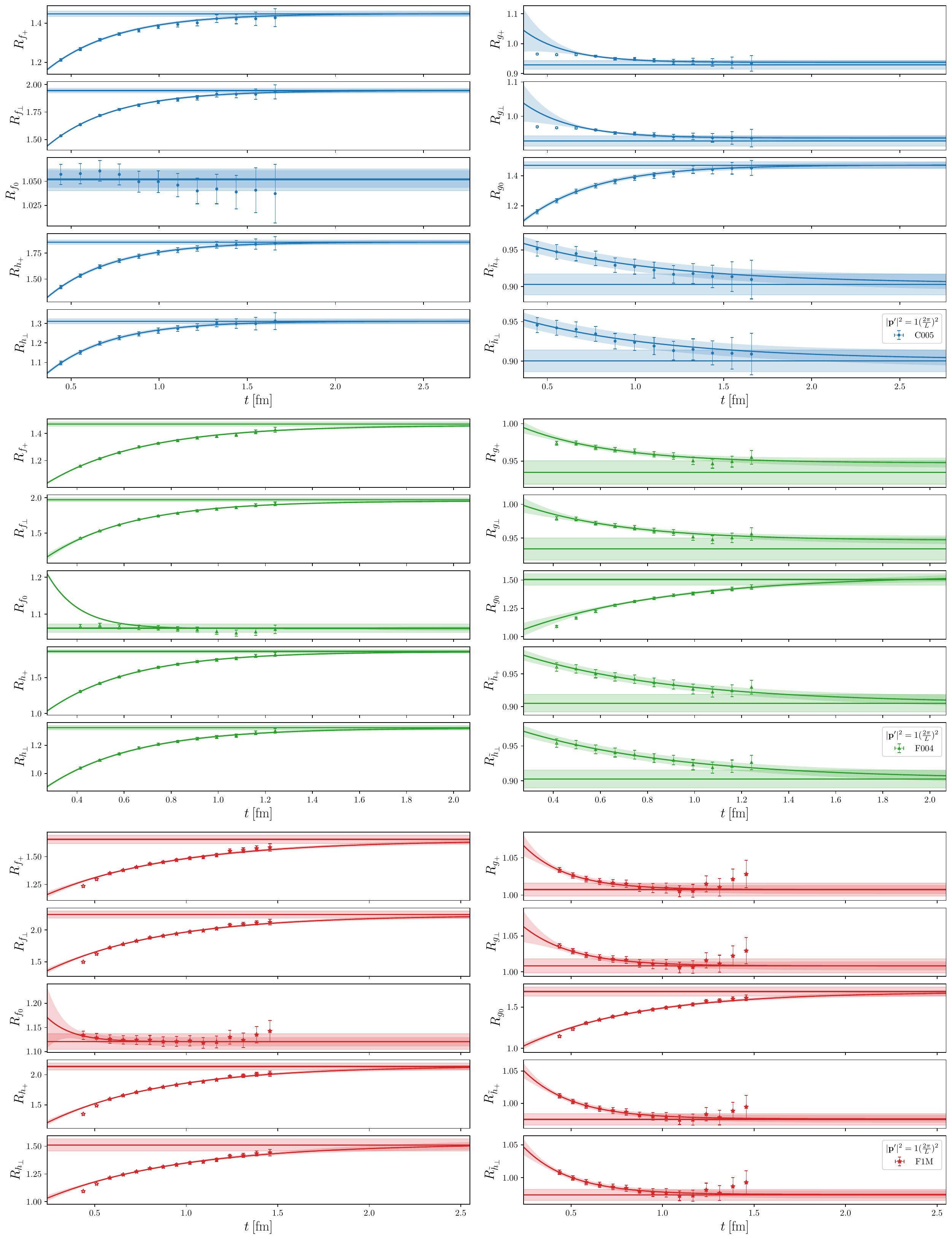}

    \caption{Like Fig.~\ref{fig:c01_ratios}, but for the other ensembles.}
    \label{fig:other_ensemble_ratios}
\end{figure}

\FloatBarrier

\section{Two-state fits of two-point functions}

In this section, we present the results of two-state fits to the baryon two-point functions of the form
\begin{equation}
C^{(2)}(t) = A (e^{-E_0 t} + B e^{-(E_0+\Delta E) t})
\end{equation}
with parameters $A$, $B$, $aE_0$, and $a\Delta E$. The $t_{\rm min}$ values, the extracted ground-state energies, and the $\chi^2/{\rm dof}$ for each fit are given in Table~\ref{tab:eff_mass_params}. Figure \ref{fig:eff_mass} shows the effective energies
\begin{equation}
a E_{\rm eff}(t+a/2) = \ln \left[C(t)/C(t+a) \right]
\end{equation}
for both the data and the model. The good $\chi^2/{\rm dof}$ values and the agreement between the effective energies of the data and those reconstructed from the fits indicate that the two-state form is sufficient for the $t_{\rm min}/a$ values used here.

\label{sec:effmass}
\begin{figure}
    \centering
    \includegraphics[width=.95\linewidth]{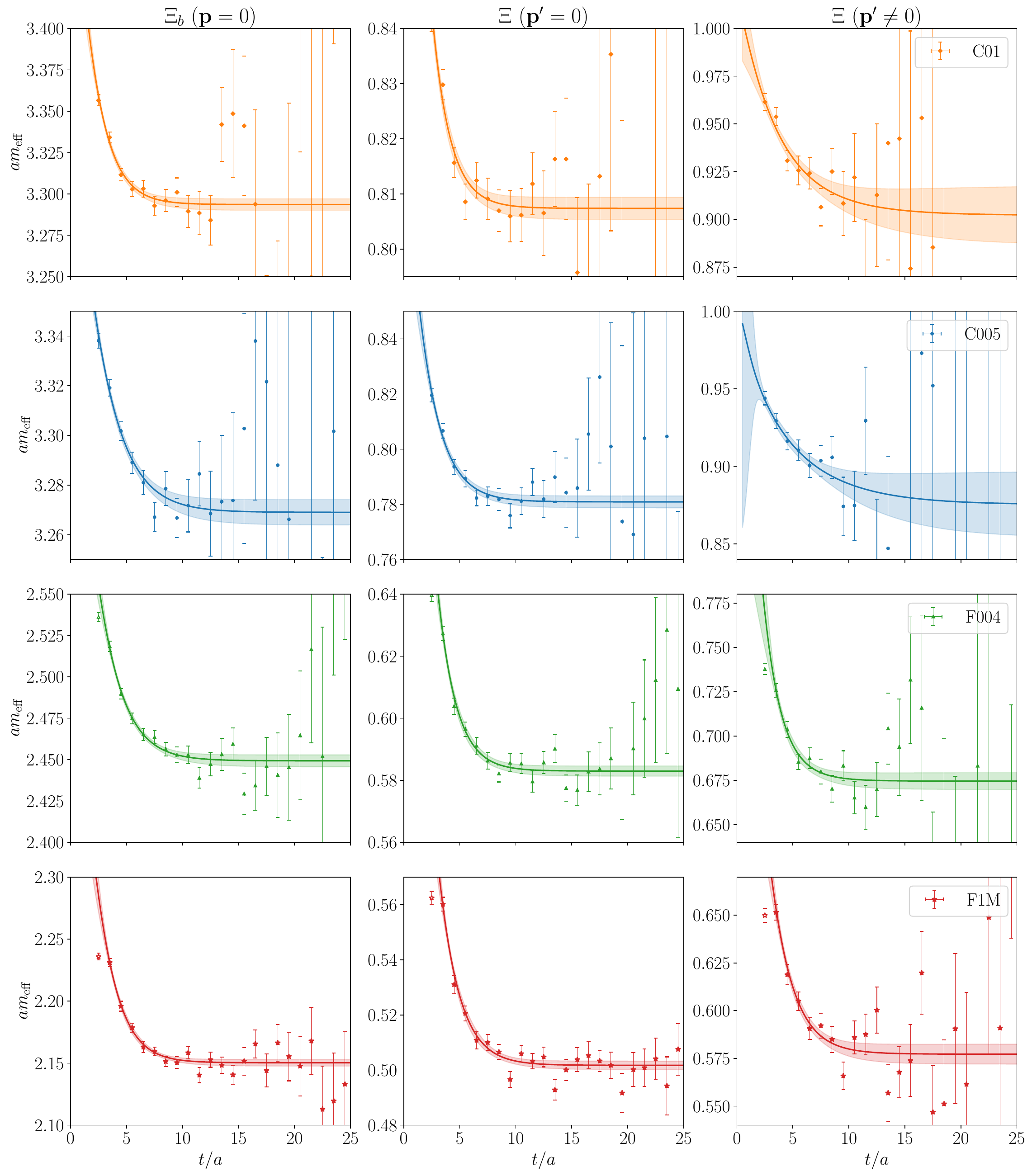}

    \caption{Effective-energy plots for both the $\Xi_b$ and $\Xi$ two-point functions at zero momentum, and for the $\Xi$ two-point functions at $|\mathbf{p}^\prime|^2=3(2\pi/L)^2$ on the C005, C01, F004 ensembles and $|\mathbf{p}^\prime|^2=5(2\pi/L)^2$ on F1M. The curves show the effective energy reconstructed from the two-state fits in the ranges listed in Table \protect\ref{tab:eff_mass_params}.}
    \label{fig:eff_mass}
\end{figure}

\begin{table}[h]
    \begin{tabular}{ccccc}
    \hline\hline
        & Ensemble & $t_{\rm min}/a$ & $aE_0$ & $\chi^2/{\rm dof}$\\
        \hline
         $\Xi_b$ $(\mathbf{p}=0)$                     & C01  &  2  &  3.2937(36) &  1.01 \\
                                                      & C005 &  2  &  3.2672(50) &  0.56 \\
                                                      & F004 &  3  &  2.4492(33) &  0.83 \\
                                                      & F1M  &  3  &  2.1504(19) &  0.87 \\
        \hline                                                          
        $\Xi$ $(\mathbf{p'}=0)$                       & C01  &  2  &  0.8075(20) &  0.59 \\
                                                      & C005 &  2  &  0.7810(21) &  0.84 \\
                                                      & F004 &  3  &  0.5829(15) &  1.06 \\
                                                      & F1M  &  3  &  0.5017(10) &  1.01 \\
        \hline                                                          
        
         $\Xi$ $(|\mathbf{p}^\prime|^2=3(2\pi/L)^2)$  & C01  &  2  &  0.906(13)  &  0.69 \\
                                                      & C005 &  2  &  0.880(18)  &  1.05 \\
                                                      & F004 &  3  &  0.6749(44) &  1.05 \\
        \hline                                                          
        $\Xi$ $(|\mathbf{p}^\prime|^2=5(2\pi/L)^2)$   & F1M  &  3  &  0.5774(38) &  0.98 \\
    \hline\hline
    \end{tabular}
    \caption{The $t_{\rm min}$ values, extracted ground-state energies, and $\chi^2/{\rm dof}$ values from two-state fits of two-point correlation functions. For each ensemble, the value of $t_{\rm min}$ is approximately half the minimum source-sink separation used in the three-point correlation functions given in Eq~(\ref{eq:threepoint}). The ground-state energies at zero momentum obtained from these two-state fits agree within uncertainties with those listed in Table~\ref{tab:hadronmasses}, which were obtained from single-state fits at larger $t_{\rm min}$ and used in the main analysis.}
    \label{tab:eff_mass_params}
\end{table}

\FloatBarrier

\section{Outer functions and susceptibilities}
\label{sec:outerfuncs}

The outer functions are taken from Ref.~\cite{Blake:2022vfl} (with the appropriate replacements of the baryon masses), and are shown here in our different notation for convenience. Following Refs.~\cite{Boyd:1997qw,Caprini:2019osi}, the outer functions are compactly expressed as
\begin{align}
    \label{eq:general-outerfunc}
        \phi_{f}(q^2)
            & =
                \frac{(m_{\Xi_b} + m_{\Xi})^{A_f} (m_{\Xi_b} - m_{\Xi})^{B_f}}{\sqrt{(16+8\cdot C_f)\cdot D_f \cdot \pi^2\cdot \chi_f}} \,
                \phi_{f,1}(q^2)^{E_f/4} \phi_{f,2}(q^2)^{F_f/4} \phi_{f,3}(q^2)^{(N_f+G_f)/2} \phi_{f,4}(q^2) 
\end{align}
with
\begin{align}
    \phi_{f,1}(q^2) &= \frac{s_-(q^2)}{z(q^2, q^2_{\rm max},t^f_+)} \, , \\
    \phi_{f,2}(q^2) &= s_+(q^2) \, ,\\ 
    \phi_{f,3}(q^2) &=  -\frac{z(q^2, 0,t^f_+)}{q^2} \, , \\ 
    \phi_{f,4}(q^2) &= 2\sqrt{t_+^f - q^2_{\rm max}} \left[1+z(q^2,q^2_{\rm max},t^f_+)\right]^{1/2}\left[1-z(q^2,q^2_{\rm max},t^f_+)\right]^{-3/2} \, . 
\end{align}
The values of $N_f$, $\chi_f$, and $A_f$ through $G_f$ are listed in Table \ref{tab:OuterFunction}. Above, $s_\pm(q^2) =(m_{\Xi_b} \pm m_\Xi)^2 -q^2$.

\begin{table}[h]
    \begin{tabular}{cccccccccc}
    \hline\hline
        $f$  & $A_f$ &  $B_f$  & $C_f$ & $D_f$ & $E_f$ & $F_f$ & $G_f$ & $N_f$ & $\chi_f$ \\
        \hline
        $f_0$ & 0 & 1 & 0 & 1 & 1 & 3 & 3 & 1 & 0.0142 \\ 
        
        $f_+$ & 1  & 0 & 1 & 2 & 3 & 1 & 3 & 2 & $0.0120/m_b^2$ \\
        
        $f_\perp$  & 0  & 0 & 1 & 1 & 3 & 1 & 2 & 2 & $0.0120/m_b^2$  \\
        
        $g_0$  & 1  & 0 & 1 & $\tfrac{2}{3}$ & 3 & 1 & 3 & 1 & 0.0157 \\
        
        $g_+$  & 0 & 1 & 0 & 3 & 1 & 3 & 3 & 2 & $0.0113/m_b^2$  \\ 
        
        $g_\perp$  & 0  & 0 & 1 & 1 & 1 & 3 & 2 & 2 & $0.0113/m_b^2$ \\
        
        $h_+$ & 0  & 0 & 1 & 2 & 3 & 1 & 1 & 3 & $0.00803/m_b^2$ \\
        
        $h_\perp$  & 1  & 0 & 1 & 1 & 3 & 1 & 2 & 3 & $0.00803/m_b^2$ \\
        
        $\tilde{h}_+$ & 0  & 0 & 1 & 2 & 1 & 3 & 1 & 3 & $0.00748/m_b^2$ \\
        
        $\tilde{h}_\perp$  & 0  & 1 & 1 & 1 & 1 & 3 & 2 & 3 & $0.00748/m_b^2$ \\
        \hline\hline
    \end{tabular}
    \caption{Parameters in the outer functions \cite{Blake:2022vfl}. Here, $m_b=4.2$ GeV.}
    \label{tab:OuterFunction}
\end{table}

\section{Asymptotic-behavior constraints}
\label{sec:asymptotic}

We define
\begin{eqnarray}
s_0^f &=& \sum_{n=0}^N a_n^f, \\
s_1^f &=& \sum_{n=1}^N n\,a_n^f, \\
s_2^f &=& \sum_{n=2}^N n(n-1)\,a_n^f, \\
s_3^f &=& \sum_{n=3}^N n(n-1)(n-2)\,a_n^f.
\end{eqnarray}
For the form factors $f_+$, $g_+$, $h_\perp$, and $\tilde{h}_\perp$, we impose the sum rules
\begin{eqnarray}
s_0^f + \sum_{n=N+1}^{N+4} a_n^f &=&0, \\
s_1^f + \sum_{n=N+1}^{N+4} n\,a_n^f &=&0, \\
s_2^f + \sum_{n=N+1}^{N+4} n(n-1)\,a_n^f &=&0, \\
s_3^f + \sum_{n=N+1}^{N+4} n(n-1)(n-2)\,a_n^f &=&0,
\end{eqnarray}
which we solve for the four coefficients $a_{N+1}^f,a_{N+2}^f,a_{N+3}^f,a_{N+4}^f$. Writing
\begin{equation}
j=N+1,
\end{equation}
we obtain
\begin{eqnarray}
 a_j^f &=& -\frac{1}{6} (j+1) (j+2) (j+3) s_0^f + \frac{1}{2} (j+1) (j+2) s_1^f - \frac{1}{2} (j+1) s_2^f + \frac{1}{6} s_3^f, \label{eq:M4j} \\
 a_{j+1}^f &=& \frac{1}{2} j (j+2) (j+3) s_0^f -\frac{1}{2} (j+2) (3 j+1) s_1^f +  \frac{1}{2}(3j+2) s_2^f -\frac{1}{2} s_3^f, \\
 a_{j+2}^f &=& -\frac{1}{2} j (j+1) (j+3) s_0^f + \frac{1}{2} j (3 j+5) s_1^f - \frac{1}{2} (3 j+1) s_2^f + \frac{1}{2} s_3^f, \\
 a_{j+3}^f &=& \frac{1}{6} j (j+1) (j+2) s_0^f -\frac{1}{2} j (j+1) s_1^f + \frac{j}{2} s_2^f -\frac{1}{6} s_3^f. \label{eq:M4jp3}
\end{eqnarray}

For the form factors $f_0$, $f_\perp$, $g_0$, $g_\perp$, $h_+$, and $\tilde{h}_+$, we impose the sum rules
\begin{eqnarray}
s_0^f + \sum_{n=N+1}^{N+3} a_n^f &=&0, \\
s_1^f + \sum_{n=N+1}^{N+3} n\,a_n^f &=&0, \\
s_2^f + \sum_{n=N+1}^{N+3} n(n-1)\,a_n^f &=&0,
\end{eqnarray}
which we solve for the three coefficients $a_{N+1}^f,a_{N+2}^f,a_{N+3}^f$. Writing again $j=N+1$, the solution in this case is given by
\begin{eqnarray}
 a_j^f &=&-\frac{1}{2} (j+1) (j+2) s_0^f+(j+1) s_1^f-\frac{1}{2}s_2^f, \label{eq:M3j} \\
 a_{j+1}^f &=& j (j+2) s_0^f-(2 j+1) s_1^f+s_2^f, \\
 a_{j+2}^f&=& -\frac{1}{2} j (j+1) s_0^f+j s_1^f-\frac{1}{2}s_2^f. \label{eq:M3jp2}
\end{eqnarray}

\section{Heavy-quark discretization errors}
\label{sec:HQdisc}

Because the three parameters of the $b$-quark action were tuned nonperturbatively using the scheme of Ref.~\cite{RBC:2012pds}, heavy-quark discretization errors from the action start at order $a^2$ and arise at tree level from two dimension-6 bilinear operators in the Symanzik effective-theory description \cite{Oktay:2008ex,Christ:2014uea}. The errors from each of the two operators are expected to be approximately of size 
\begin{equation}
f_E(am_Q,\nu,c_{E,B})(a \Lambda_{\rm had})^2, \label{eq:fEerrors}
\end{equation}
where the ``mismatch function'' $f_E$ is given in Eq.~(B1) of Ref.~\cite{Christ:2014uea} (that reference uses the notation $am_Q=am_0$, $\zeta=\nu$, $c_P=c_{E,B}$). The numerical values of $f_E$ and $f_E\cdot(a \Lambda_{\rm had})^2$ for our choice of $b$-quark parameters \cite{Meinel:2023wyg} are given in Tables \ref{tab:mismatch} and \ref{tab:HQerrors}, respectively. We find that $f_E$ is nearly constant, which means that our continuum extrapolation can nearly perfectly remove these errors. In addition, discretization errors arise from the $b\to s $ currents. For the vector and axial-vector currents, which include full one-loop order-$a$ improvement, the leading discretization errors are of order $a^2$ (due to missing dimension-6 corrections to the currents) and $\alpha_s^2 a$ (due to missing two-loop corrections to the coefficients of the dimension-5 operators). The uncertainties due to the missing two-loop corrections can be estimated through the scale dependence of the one-loop corrections and are already included in our analysis through the random sampling of the improvement coefficients. The mismatch functions of the three dimension-6 operators in the $b\to s$ currents, denoted as $f_{X_1}$, $f_{X_2}$, and $f_Y$, are given in Eqs.~(B9)-(B11) of Ref.~\cite{Christ:2014uea} (based on Ref.~\cite{El-Khadra:1996wdx}), and the corresponding numerical values are also given in Tables \ref{tab:mismatch} and \ref{tab:HQerrors}. These mismatch functions show more dependence on the heavy-quark parameters, but the products with $(a \Lambda_{\rm had})^2$ are still seen to decrease with the lattice spacing and are fairly small. This indicates that our continuum extrapolation will still partially remove these errors, and any remaining effect should be negligible compared to the overall uncertainties. For the $b\to s$ tensor currents, which were order-$a$-improved at tree level only, there are also discretization errors proportional to $\alpha_s a$. However, based on a comparison of tree-level and one-loop order-$a$ improvement for the vector and axial-vector form factors, we expect the size of the $\alpha_s a$ corrections to be $\lesssim1\%$, and hence negligible compared to the $\approx 5.3\%$ estimate of the tensor-current matching uncertainty included in the results.

\begin{table}
\begin{tabular}{cccccc}
\hline\hline
Ensemble & $f_E$ & $f_{X_1}$ & $f_{X_2}$ & $f_{Y}$ \\ 
\hline
C01, C005 & 0.0589 & 0.0947 & 0.1479 & 0.1243 \\ 
F004      & 0.0554 & 0.1200 & 0.1630 & 0.1023 \\ 
F1M       & 0.0529 & 0.1353 & 0.1791 & 0.0970 \\ 
\hline\hline
\end{tabular}
\caption{\label{tab:mismatch}Values of the heavy-quark mismatch functions.}
\end{table}

\begin{table}
\begin{tabular}{cccccc}
\hline\hline
Ensemble & $f_E\cdot(a \Lambda_{\rm had})^2$ & $f_{X_1}\cdot(a \Lambda_{\rm had})^2$ & $f_{X_2}\cdot(a \Lambda_{\rm had})^2$ & $f_{Y}\cdot(a \Lambda_{\rm had})^2$ \\ 
\hline
C01, C005 & 0.17\% & 0.27\% & 0.42\% & 0.35\% \\ 
F004      & 0.09\% & 0.19\% & 0.26\% & 0.16\% \\ 
F1M       & 0.06\% & 0.17\% & 0.22\% & 0.12\% \\ 
\hline\hline
\end{tabular}
\caption{\label{tab:HQerrors}Values of the heavy-quark mismatch functions times $(a \Lambda_{\rm had})^2$.}
\end{table}

\FloatBarrier
\providecommand{\href}[2]{#2}\begingroup\raggedright\endgroup

\end{document}